\documentclass[twocolumn]{aastex631}

\newcommand{\Msun}{${\rm M}_{\odot}$}

\newcommand{\kms}{km\,s$^{-1}$}

\newcommand{\hi}{\ion{H}{1}}

\newcommand{\cii}{\ion{C}{2}}

\newcommand{\civ}{\ion{C}{4}}

\newcommand{\siii}{\ion{Si}{2}}
\newcommand{\siiii}{\ion{Si}{3}}
\newcommand{\siiv}{\ion{Si}{4}}

\newcommand{\mgii}{\ion{Mg}{2}}

\newcommand{\ovi}{\ion{O}{6}}

\usepackage{graphicx}
\usepackage{graphics}
\usepackage{amsmath}
\usepackage{amssymb}
\usepackage{url}
\usepackage{hyperref}
\usepackage{xcolor}
\usepackage{physics}
\usepackage{rotating}

\submitjournal{\apj}
\shorttitle{FOGGIE Clouds}
\shortauthors{Augustin et al.}

\begin{document}

\title{FOGGIE X: Characterizing the Small-Scale Structure of the CGM and its Imprint on Observables}

\author[0000-0001-7472-3824]{Ramona Augustin}
\affiliation{Leibniz-Institut f{\"u}r Astrophysik Potsdam (AIP), An der Sternwarte 16, 14482 Potsdam, Germany}
\correspondingauthor{Ramona Augustin}
\email{raugustin@aip.de}
\affiliation{Space Telescope Science Institute, 3700 San Martin Dr., Baltimore, MD 21218}

\author[0000-0002-7982-412X]{Jason Tumlinson}
\affiliation{Space Telescope Science Institute, 3700 San Martin Dr., Baltimore, MD 21218}
\affiliation{Center for Astrophysical Sciences, William H.\ Miller III Department of Physics \& Astronomy, Johns Hopkins University, 3400 N.\ Charles Street, Baltimore, MD 21218}

\author[0000-0003-1455-8788]{Molly S.\ Peeples}
\affiliation{Space Telescope Science Institute, 3700 San Martin Dr., Baltimore, MD 21218}
\affiliation{Center for Astrophysical Sciences, William H.\ Miller III Department of Physics \& Astronomy, Johns Hopkins University, 3400 N.\ Charles Street, Baltimore, MD 21218}

\author[0000-0002-2786-0348]{Brian W.\ O'Shea}
\affiliation{Department of Computational Mathematics, Science, \& Engineering, Michigan State University, 428 S. Shaw Lane, East Lansing, MI 48824}
\affiliation{Department of Physics \& Astronomy, 567 Wilson Road, Michigan State University, East Lansing, MI 48824}
\affiliation{Facility for Rare Isotope Beams, Michigan State University, 640 S. Shaw Lane, East Lansing, MI 48824}
\affiliation{Institute for Cyber-Enabled Research, 567 Wilson Road, Michigan State University, East Lansing, MI 48824}

\author[0000-0002-6804-630X]{Britton D.\ Smith}
\affiliation{Institute for Astronomy, University of Edinburgh, Royal Observatory, EH9 3HJ, UK}

\author[0000-0003-1785-8022]{Cassandra Lochhaas}
\affiliation{Space Telescope Science Institute, 3700 San Martin Dr., Baltimore, MD 21218}
\affiliation{Center for Astrophysics, Harvard \& Smithsonian, 60 Garden St., Cambridge, MA 02138}
\affiliation{NASA Hubble Fellow}

\author[0000-0002-1685-5818]{Anna C.\ Wright}
\affiliation{Center for Astrophysical Sciences, William H.\ Miller III Department of Physics \& Astronomy, Johns Hopkins University, 3400 N.\ Charles Street, Baltimore, MD 21218}
\affiliation{Center for Computational Astrophysics, Flatiron Institute, 162 Fifth Avenue, New York, NY 10010}

\author[0000-0003-4804-7142]{Ayan Acharyya}
\affiliation{Center for Astrophysical Sciences, William H.\ Miller III Department of Physics \& Astronomy, Johns Hopkins University, 3400 N.\ Charles Street, Baltimore, MD 21218}
\affiliation{INAF - Astronomical Observatory of Padova, vicolo dell’Osservatorio 5, IT-35122 Padova, Italy}

\author[0000-0002-0355-0134]{Jessica K.\ Werk}
\affil{University of Washington, Department of Astronomy, Seattle, WA 98195, USA}

\author[0000-0001-9158-0829]{Nicolas Lehner}
\affiliation{Department of Physics and Astronomy, University of Notre Dame, Notre Dame, IN 46556}

\author[0000-0002-0646-1540]{Lauren Corlies}
\affiliation{University of California Observatories/Lick Observatory, Mount Hamilton, CA 95140, USA}

\author[0000-0002-6386-7299]{Raymond C.\ Simons}
\affiliation{Department of Engineering and Physics, Providence College, 1 Cunningham Sq, Providence, RI 02918 USA}

\author[0000-0002-2591-3792]{J.\ Christopher Howk}
\affiliation{Department of Physics and Astronomy, University of Notre Dame, Notre Dame, IN 46556}

\author[0000-0002-7893-1054]{John M.\ O'Meara}
\affiliation{W.\ M.\ Keck Observatory, Waimea, HI 96743}

\begin{abstract}
One of the main unknowns in galaxy evolution is how gas flows into and out of galaxies in the circumgalactic medium (CGM). Studies observing the CGM in absorption using multiple or extended background objects suggest a high degree of  variation on relatively small ($\lesssim 1$\,kpc) spatial scales. Similarly, high-resolution simulations generally exhibit small-scale substructure in the gas around galaxies. We examine the small-scale structure of the $z = 1$ CGM using simulations from the FOGGIE (Figuring Out Gas \& Galaxies in Enzo) project. We select gaseous substructures (``clumps'')  by their local overdensity and investigate their physical properties, including  temperature, metallicity, and kinematics with respect to the galaxy and the nearby surroundings. FOGGIE resolves clumps down to sphericalized radii $R \sim 0.25$ kpc at $z = 1$. The distribution of clumps peaks at $\sim 10^5$ \Msun\ and $10^{4}$\,K, consistent with relatively condensed, cool gas with a slight preference for inflow-like velocities. Many clumps show internal temperature and density variations, and thus internally varying ionization levels for key diagnostic ions such as \ion{H}{1}, \ion{Mg}{2}, and \ion{O}{6}. The average metallicity in clumps is about a factor 1.5--2$\times$ lower in metallicity than nearby gas, suggesting that the metals are not well-mixed between structured and diffuse CGM, which may have implications for observational metallicity estimations of dense CGM clouds. 
We estimate the survivability of CGM clumps and find that structures larger than 0.5 kpc are generally long-lived.
Finally, we qualitatively compare the simulated cloud properties to Milky Way high-velocity clouds.  
\end{abstract}

\section{Introduction}
The gas surrounding galaxies (the circumgalactic medium, or CGM) acts as a mediator of galactic star formation, feedback, and recycling \citep*{Tumlinson2017}. The emergent ``baryon cycle'' helps to set galactic masses and star formation rates and may play a role in galaxy quenching. The available observational information about the CGM rests primarily on absorption-line studies from the classical quasar absorber technique or, more recently, on resolved absorption or emission from integral-field spectrographs (IFS). Both techniques are biased. Absorption studies generally convey no spatial information about the absorbing gas structures, which are seen in projection and only along a narrow ``pencil beam'' sightline, while emission-line techniques preferentially see the denser gas (emissivity scales as density-squared). 
{Both the single-sightline absorption observations (due to limited sampling of the halo) and the emission observations (due to typically poor spatial resolution and bias on the highest density regions) provide little to no information on the spatial variation of physical properties such as density, temperature and metallicity on small scales ($\lesssim 10$ kpc). }
Therefore, physical inferences about the mass, metal content, or dynamics of the CGM make assumptions about cloud sizes and density of the gas to determine average CGM properties (e.g., \citealp{Werk2014, Zahedy2019, Qu23}). 

We are motivated to investigate the small-scale structure of the CGM from four different lines of observational evidence, both direct and indirect. The first line of evidence is that absorption lines vary in strength from halo to halo. Even though single absorption sightlines do not individually probe the small scale structure of the CGM, surveys using one sightline per galaxy for a sample of galaxies find strong indirect evidence for an inhomogeneous CGM with a wide range of density, temperature, and ionization state. Absorption lines from \ion{H}{1} \citep{Tumlinson2013}, \ion{Mg}{2} \citep{Nielsen2013, Zahedy2016}, and \ion{C}{4} \citep{Bordoloi2014} vary across orders of magnitude in samples of $\sim 10$ or more galaxies, indicating that the typical CGM has a similar degree of internal variation. The high ion \ion{O}{6} varies less from halo to halo \citep{Tumlinson2011}, which has been interpreted as evidence that the higher ionization gas occupies a more volume-filling medium with fewer clumps \citep{Werk2016}. From these patterns we can infer that the CGM is structured in density and temperature on $\lesssim 10$ kpc scales, probably with smaller structures in cooler, lower-ionization gas. Another line of complementary evidence for small-scale structure from single-sightlines is large metallicity variations (from a factors of a few to $>100\times$ difference) observed in absorbers that are separated by $\delta v < 500$\,\kms\ \citep{Lehner2019,Lehner2022}.

\begin{figure*}[!t]
    \centering
    \includegraphics[width=.99\textwidth]{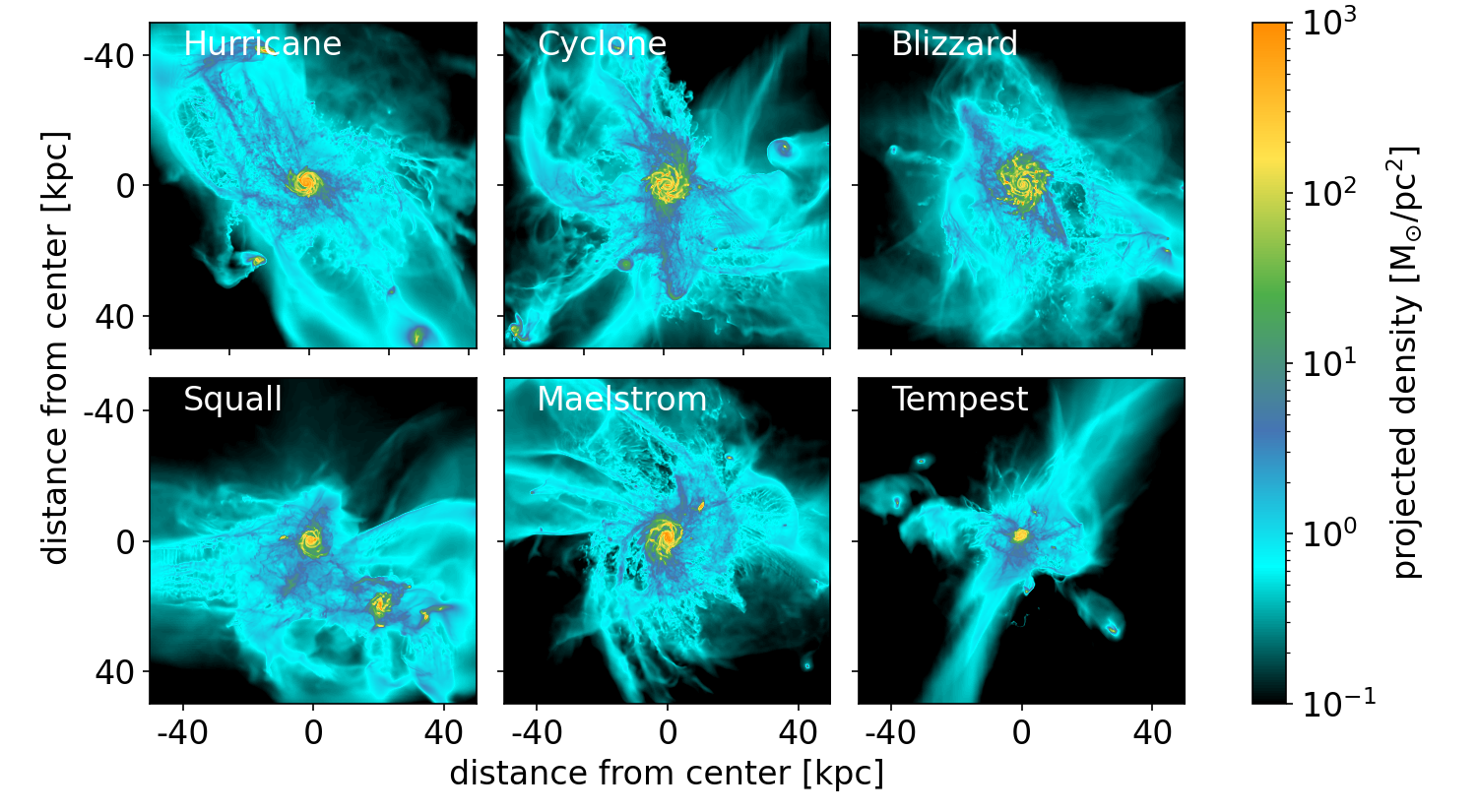}
    \caption{Gas density projection maps of the six FOGGIE halos at $z = 1$. These density projections clearly show clumpy structures throughout the gas halos, concentrated within the inner 40 kpc of the CGM.}
    \label{fig:halosatz1}
\end{figure*}

A second factor motivating small-scale investigations of the CGM comes from spatially-resolved absorption and emission studies. Observations of absorbers toward lensed quasars have long been used to directly constrain the sizes of absorbers \citep{Rauch2002}, as multiply-lensed QSOs can provide sightlines within 1 kpc of each other.  Additionally, studies using lensed and unlensed quasars have detected significant amounts of cool gas around massive elliptical quenched galaxies \citep{Thom2012, Zahedy2016, Chen2018} that appears more patchy than around star-forming galaxies, suggesting differences in the small scale structure of the CGM.

Integral field units (IFUs) have enabled 2-D maps of absorbers seen against extended \citep[e.g.,][]{Peroux2018} or lensed galaxies \citep[e.g.][]{Lopez2018,Bordoloi2022}. Size constraints also come from hit rates and absorption variation in the separate sight-lines \citep[e.g.,][]{Lan2017,Rubin2018,Augustin2021}. These studies indicate that there is significant variation in the absorption strength over the angular extent of the background source, which can span kpc to tens of kpc scales. This variation within single systems indicates that there is an underlying variation in spatial structure of the gas in terms of density, ionization, and/or metallicity that is not subject to the ambiguity of aggregating multiple sightlines. 

Mapping emission from circumgalactic gas has only recently become feasible, again through the use of IFUs \citep{Wisotzki2016, Arrigoni2019,Guo2024}. However, such observations remain challenging and are mostly achievable only for the brightest emission line, Ly$\alpha$, and in some cases for bright metal lines \citep{Burchett2021, Dutta2023, Kusakabe2024, Nielsen2024}. Because these observations are typically close to instrumental limits in surface brightness they typically do not provide constraints on the morphology of the gas clouds in the CGM, particularly in the resonant lines like Ly$\alpha$, and generally not at spatial scales down to $\sim 1$ kpc. For a few bright or large extended sources, full 2-D maps are possible and show clear substructure at the $\lesssim$ 10 kpc scales \citep{Rupke2019}. 

A third major indication of a highly structured CGM comes from the few cases of nearby galaxies for which we have multiple sightlines through the CGM \citep[e.g.,][]{Keeney2013,Bowen2016}. The halo of the Milky Way's near neighbor, M31, subtends a larger angle on the sky than any other galaxy of similar mass, providing a unique case for which many pencil beam absorber sightlines can be observed. Project AMIGA \citep{Lehner2020} surveyed 43 quasar sightlines passing within 25--569 kpc of the M31 disk, using all the key UV ions accessible to the Cosmic Origins Spectrograph (COS) on the {\em Hubble Space Telescope}. By coincidence, this is the same number of sightlines for which the COS-Halos program collected one sightline per background galaxy \citep{Tumlinson2013}. Project AMIGA shows that the CGM of a single $L^*$ disk galaxy qualitatively has as much scatter as a similar sample of many galaxies. Project AMIGA finds that the covering fraction of the multiphase ions (\cii--{\sc{iv}}, \siii--{\sc{iv}}, etc.) is high inside 40--50 kpc, but with a high degree of scatter reflecting varying density, temperature, and/or ionization on $\lesssim 10$ kpc scales. Beyond 50\,kpc, the CGM of M31 is more diffuse and less complex (but likely still multiphase), with high covering fractions of \siiii\ and \ovi. Since all these sightlines pass through the halo of a single galaxy, this is strong direct evidence for a structured CGM. 

A fourth and final line of evidence for small-scale structure comes from maps of the nearby gaseous halo of the Milky Way \citep*{Putman2012}. All-sky maps from radio observations ($N_{\rm HI} > 10^{18.5}$ cm$^{-2}$) show that the Milky Way halo is rich in structure at scales from $>10$\,kpc, such as the Magellanic Stream, down to sub-degree scales, and it is common for large structures in the maps to break into smaller ones when observed at higher spatial resolution \citep{Westmeier2018}. The aggregate map compiled in the review by \citeauthor*{Putman2012} shows hundreds of Compact High Velocity Clouds (CHVCs) with angular sizes around a degree or less. While the angular sizes and relative velocities of these CHVCs are easily measured in radio data, their metallicities, temperatures, and pressures are mostly unconstrained as a population (aside from a few individual cases; see, e.g., \citealt{Ashley2024}) and so their origins are poorly understood. If their physical sizes are 1--2 kpc, as implied by their angular size and likely distances, they may contribute a non-negligible fraction of the Milky Way's gaseous fuel \citep*{Putman2012}. In light of this evidence, we are motivated to explore small-scale structure in simulations to better understand the physics that creates these CHVCs.

These observational findings strongly suggest that the CGM has spatial structure on scales below 1--10\,kpc, but do not answer directly how these small scale structures arise from or influence galactic accretion and feedback. From a theoretical point of view, there are strong reasons to believe that small-scale structure both influences and are governed by feedback and the accretion of gas in the CGM \citep{CAFG2023}. Small-scale clouds can be the direct consequence of gas cooling instabilities \citep[``precipitation'']{Voit2019b} or result from the fragmentation of larger clouds or filaments from the IGM. Small-scale structure forms a part of most general models of galaxy accretion (\citealp{Keres2005,Keres2009,Brooks2009,Stern2019,Hafen2022}), as either the primary channel for inflows themselves or the cooled instabilities from hotter gas that follow a ``hot mode'' scenario. Recently, cosmological hydrodynamic simulations have become able to simulate CGM gas at high resolution, 
with spatial and mass resolution better than 1\,kpc and $\sim 100~M_{\odot}$, respectively. These simulations naturally yield a clumpy, dynamic CGM owing to the joint effects of cooling, turbulence, and/or energetic feedback \citep{vandeVoort2019, Peeples2019, Hummels2019, Suresh2019,Ramesh2024}. At the bottom end of the spatial scale, simulations of gas turbulence and cooling in idealized situations or small patches of CGM suggest that physically coherent, identifiable structures may continue to form from cooling instabilities and fragmentation to, or below, the parsec scale \citep{McCourt2018, Gronke2022}, a regime not (yet) accessible to cosmological zooms but which nevertheless might be the ultimate endpoint of evolution for real CGM gas. 

All these observational and theoretical considerations taken together leave us with several questions about the CGM: How is star-forming fuel spatially distributed around normal galaxies? How does small-scale structure reflect the governing physics? How does cool gas survive in the otherwise hot halo around a massive galaxy? How does it make it onto the galaxy and feed star formation? Are small-scale clouds a significant source of galactic accretion? 
Finding answers to these questions will ultimately bring us to a better understanding of galaxy evolution and formation at large.
For the moment, however, a first step toward these answers is to develop a general understanding for CGM small scale structure. 
The FOGGIE simulations, due to their high spatial resolution in the CGM of six Milky-Way progenitor galaxies, make a perfect testbed for characterizing small-scale structures in the CGM.
In this work we will extract overdensities from the halos of these simulated galaxies, study their physical properties, and compare them to their diffuse surroundings.

This paper is structured as follows. Section~\ref{sec:simulations} describes the FOGGIE simulations and our clump-oriented analysis methods. 
Section~\ref{sec:global} describes the global distributions of clump properties. 
Section~\ref{sec:individual_clumps} zooms in to investigate the properties of some example clumps with notable internal structure and other interesting properties. 
Section~\ref{sec:discussion} makes comparison to small-scale structure seen in observations and in other simulations. 
Section \ref{sec:conclusions} reviews our results and draws larger conclusions. 

Our simulations use the following cosmology: $\Lambda$CDM \citep{Planck:2014aa}, $1 - \Omega_\Lambda = \Omega_m = 0.285$, $\Omega_b = 0.0461$, and $h=0.695$.
Throughout the paper, distances are denoted in physical units unless otherwise specified.

\section{The Simulations and Clump Extraction}
\label{sec:simulations}

We describe the FOGGIE simulations analyzed throughout this paper in Section~\ref{sec:sims}, and in Section~\ref{sec:id} we describe how we identify the small-scale ``clumps'' that are our basis for describing the small-scale structure of the circumgalactic medium.

\subsection{Simulations}\label{sec:sims}
For the present analysis we use the FOGGIE (Figuring Out Gas \& Galaxies In Enzo) simulations, presented in detail in earlier works \citep{Peeples2019, Corlies2020, Zheng2020, Simons2020, Lochhaas2021, Lochhaas2023, Wright2024, Acharyya2024, Simons2024}. These simulations were designed to resolve the CGM at a level not achieved previously in cosmological zooms, so they are particularly well-suited to study the small-scale structure of the CGM. FOGGIE was run with the Enzo Adaptive Mesh Refinement (AMR) code \citep{Bryan2014,BrummelSmith2019}, which enables high spatial resolution in diffuse gas. FOGGIE's production runs refine down to gas cells as small as 137 pc at $z = 1$. We are therefore able to trace even small variations in spatial distribution of the CGM gas. 
Because there is no restriction on minimum gas {\em mass} per resolution element in AMR, the median mass of cells in the CGM of the halos analyzed here is $\sim$ 44\,M$_{\odot}$ and some cells reach as low as $\lesssim 1\, \rm M_{\odot}$ \cite[see also][]{Lochhaas2023}. 

As introduced in \citet{Peeples2019}, to attain this level of refinement we use ``forced refinement,'' which tracks the targeted galaxy with a cubic volume 200 kpc$/h$ comoving on a side, centered around the moving galactic center of mass. Within this ``forced refinement box'' cells have a minimum size of 548\,pc at $z=1$. We additionally enforce refinement within this region in cells where the cooling time is shorter than the free fall time, down to 137\,pc. This ``cooling refinement'' ensures that additional resolution is placed where it is most needed, and in these runs a large fraction (90--99\%) of the CGM mass is resolved according to the cooling time criterion \citep{Simons2020}.

\begin{figure}
    \centering
    \includegraphics[width=.5\textwidth]{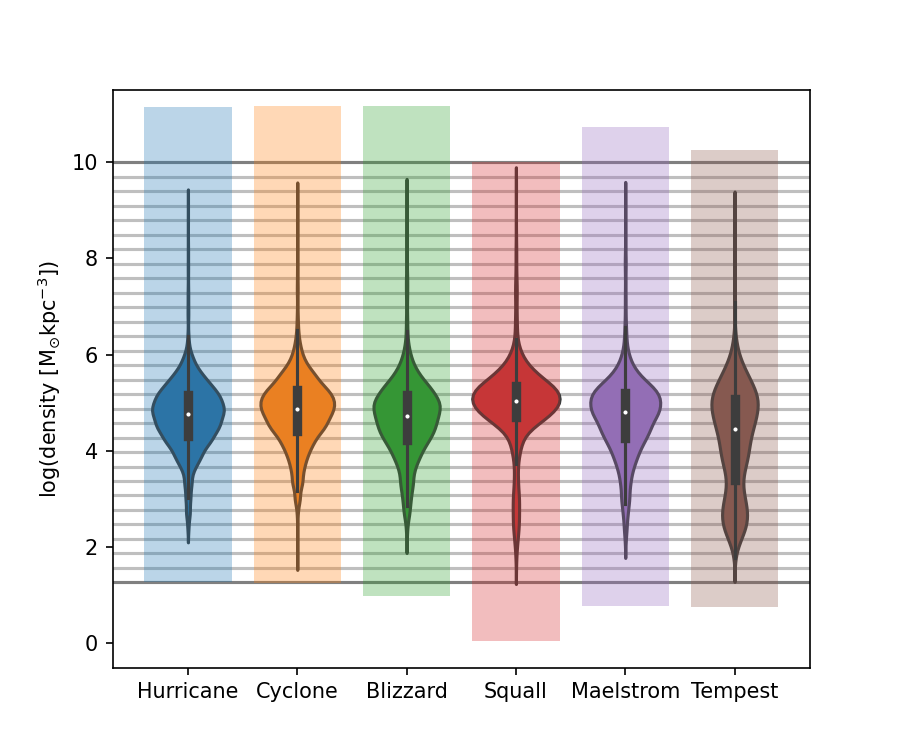}
    \caption{Density spread within each of the six FOGGIE halos (colored bars in the background) as well as the chosen minimum detection thresholds when identifying clumps (grey horizontal lines). The violin plots indicate the distributions of densities {found in the cells} of the identified clumps in each halo. The global minimum threshold is set by Hurricane and the global maximum by Squall.}
    \label{fig:denscuts}
\end{figure}

The FOGGIE simulations implement simple gas physics, including radiative cooling with a self-shielding approximation for high density gas, star formation, and thermal feedback from SNe. 
{We note that while it has been demonstrated that thermal feedback can lead to numerical overcooling in gas cells where the cooling time is shorter than the freefall time, particularly in groups and clusters of galaxies (see, e.g., \citealt{McCourt2012,Voit2019b}), the high resolution of our simulations mitigates this problem somewhat \citep[see, e.g.,][]{Hummels2019}, particularly since we focus on clouds in the CGM where the cooling length is typically resolved.
However, due to overcooling in the ISM, our simulations create more stars than they should, causing our galaxies do divert from the common stellar mass -- halo mass relation. For details about this caveat we refer the reader to \citep{Wright2024}.
}
We do not include other forces that may have significant effects on gas structure but are neither theoretically nor observationally well-constrained, such as magnetic fields, cosmic rays, or AGN feedback. 
We focus here on the abundance and physical properties of clumps in relation to the host galaxy, and leave examination of their behavior under the influence of these other effects to future work. 

\begin{figure*}
    \centering
    \includegraphics[width=.99\textwidth]{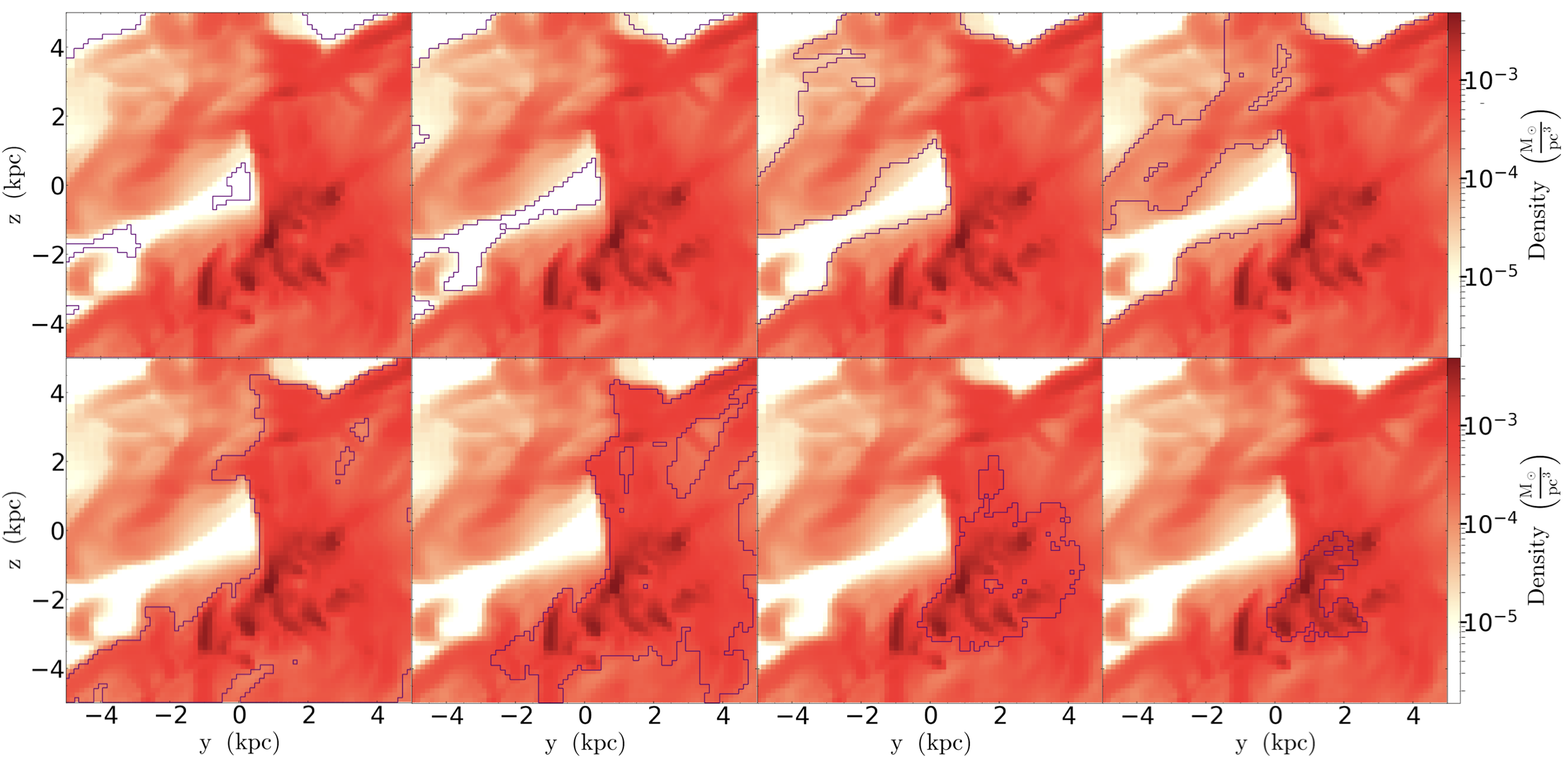}
    \caption{Illustration of the clump-finding process (from upper left to lower right). Starting at lower density isocontours, coherent structures are identified and within them a higher density threshold is applied to find ``children'' structures within the previous one until we arrive at a set of ``leaves'' which are the highest density structures within a certain environment.}
    \label{fig:clumpfinding}
\end{figure*}

For our analysis we use all six FOGGIE halos at $z=1$, as shown in density projections in Figure \ref{fig:halosatz1}. All these halos were selected to be Milky Way-like at $z = 0$ in terms of their mass and merger history, which is relatively quiescent after $z = 2$ (see \citealt{Wright2024} for a thorough discussion). By $z = 1$ they have already had all their major mergers and experience only relatively minor mergers thereafter\footnote{The only exception to this pattern is that Squall experiences a 4:1 merger at $z = 0.7$ with the infalling galaxy seen to its lower-right in Figure~\ref{fig:halosatz1}.}. 
{Their basic properties are summarized in Table \ref{tab:haloprops}.}
The FOGGIE halos form stars faster and show more small-scale structure at $z = 1$ than at lower redshifts; by $z = 0$ the CGM is more settled with comparatively less obvious structure remaining. {This means the FOGGIE halos at $z = 0$ show much fewer small-scale clumps than at $z = 1$.} Also, this choice of redshift enables us to tie our theoretical findings on small scale CGM structure back to observables, particularly for absorbers towards lensed quasars studied at visible wavelengths.  The \ion{Mg}{2} ion is the most common tracer at $z \sim 1$ because it is a strong doublet that is readily covered at visible wavelengths along the line of sight to high-redshift quasars \citep[e.g.,][]{Chen2018,Rubin2018,Kulkarni2019,Zahedy2019,Augustin2021}.
This will allow us to provide strong theoretical constraints on clump sizes that can be used to interpret and compare to absorber size constraints from the observational data (see Sec. \ref{sec:compareobs}).
{We emphasize here that we choose $z = 1$ for this analysis as the strongest constraints on CGM small scale structure to date come from absorber variations towards lensed objects. While clumps at $z = 0$ might appear to be a better comparison sample to the HVCs at face-value, the observed constraints on HVC sizes are typically in angular scale rather than physical scale, due to unknown distances. Therefore $z = 1$ clumps are a better choice for comparing physical sizes. Nevertheless, we include a qualitative comparison to local HVCs in Sec. \ref{sec:compareobs}.}

\begin{table*}[]
    \centering
    \begin{tabular}{c|cccccc}
       halo name  & stellar mass & gas mass & dark matter mass & total virial mass & star-formation rate & virial radius \\ \newline
       [ID number]  & [$\rm M_{\odot}$] & [$\rm M_{\odot}$] & [$\rm M_{\odot}$] & [$\rm M_{\odot}$] & [$\rm M_{\odot} yr^{-1}$] & [kpc]\\
         \hline
       Hurricane [002392]  & $\rm 1.5 \times 10^{11}$ & $\rm 4.9 \times 10^{10}$ & $\rm 8.0 \times 10^{11}$ & $\rm 10.0 \times 10^{11}$  & 11.4 & 140 \\
       Cyclone [002878] & $\rm 1.5 \times 10^{11}$ & $\rm 4.85 \times 10^{10}$ & $\rm 8.0 \times 10^{11}$ &  $\rm 10.0 \times 10^{11}$ & 25.9 & 143 \\
       Blizzard [004123] & $\rm 9.0 \times 10^{10}$ & $\rm 3.9 \times 10^{10}$ & $\rm 6.00 \times 10^{11}$ & $\rm 7.3 \times 10^{11}$  & 43.5 & 129 \\
       Squall [005016] & $\rm 4.0 \times 10^{10}$ & $\rm 3.3 \times 10^{10}$ & $\rm 3.5 \times 10^{11}$ & $\rm 4.3 \times 10^{11}$  & 4.9 & 108 \\ 
       Maelstrom [005036] & $\rm 6.9 \times 10^{10}$ & $\rm 4.3 \times 10^{10}$ & $\rm 6.1 \times 10^{11}$ & $\rm 7.3 \times 10^{11}$  & 18.8 & 129 \\
       Tempest [008508] & $\rm 3.8 \times 10^{10}$ & $\rm 1.1 \times 10^{10}$ & $\rm 2.7 \times 10^{11}$ & $\rm 3.2 \times 10^{11}$  & 6.1 & 98 \\
       \hline
    \end{tabular}
    \caption{Basic halo properties (within the virial radius) of the six halos used for this study at redshift $z = 1$.}
    \label{tab:haloprops}
\end{table*}

\subsection{Clump Identification}
\label{sec:id}

To identify CGM clumps and extract them from their ambient environment, we use the clump-finding algorithm provided by {\tt yt} \citep{Smith2009,Turk2011}. The goal is to find physical clumps without regard to temperature, metallicity, or other properties, and then to investigate those other physical and observational properties.  This allows us to assess the full range of properties that a physical clump may have without biasing ourselves towards, e.g., \ion{H}{1}-selected clumps only. 

Our clump-identification method is based on {\texttt{yt}}'s clump-finding algorithm \citep{Turk2011}, and works as follows.
We identify density isocontours {in each halo} at a given threshold in gas density. The range of threshold densities is defined by the minimum and maximum gas density across {\em all} six halo snapshots, so that cells between the two  thresholds are guaranteed to exist in every halo (Figure~\ref{fig:denscuts}). The resulting density range is from $1.3\times 10^{-30}\,\rm{g{\,}cm^{-3}}$ ($1.9\times 10^{-8}\,\rm{M_{\odot}{\,}pc^{-3}}$) to $2.1\times 10^{-22}\, \rm{g{\,}cm^{-3}}$ ($3.1\,\rm{M_{\odot}{\,}pc^{-3}}$). Starting with the lowest threshold, at each iteration the threshold density is multiplied by a factor of two and a coherent structure above the threshold density of the current iteration is identified. If the identified structure is part of a previously-identified structure, it is linked to that previous structure along its ``branch'' and we end up with a tree-like architecture of clumps in the CGM (see Fig. \ref{fig:clumpfinding} for an illustration of density isocontours at different density thresholds). 
Once there are no further substructures found within a given isocontour this structure becomes a ``leaf'' in  this tree and we define it as a ``clump'' for the purposes of our analysis.
The clumps are thus natural overdensities with respect to the ambient medium but can have any temperature, metallicity, kinematics, or other properties. In particular, the clumps can be at any {\em absolute} density so long as they are overdense relative to their environs.
Finally, we require that each clump consists of at least 20 contiguous cells to ensure they are well-resolved and contain cells in all three spatial directions.

For completeness and to explore the radial trend of the clumps in the CGM, we apply the clump finding procedure to a sphere with radius 80 kpc  around the central galaxy. The forced-refinement box at $z=1$ has a width of 143 kpc, so our identified clumps should be robust out to 71 kpc but may suffer from resolution effects at the boundaries of the refinement box farther out\footnote{To make the clump finding computationally feasible, we separately analyze four concentric spherical shells of thickness 20 kpc over at 0--80 kpc from the halo center. This size is much larger than the average clump size of $\leq$ 1 kpc, and so structures right at those boundaries are artificially cut with minimal impact to the final results.}.

\section{Properties of Clumps}
\label{sec:global}

\begin{sidewaystable}[ht]
\scriptsize
\centering
\begin{tabular}{|c|c|c|c|c|c|c|c|c|c|c|c|c|c|c|c|}
\hline
 Clump & Halo  & Mass & Volume & Size & Elon- & radial & Distance from & Number & Metallicity & Pressure & Env. & Env. & Env. & ISM & Satellite\\
 ID & ID  &  &  &  & gation & velocity & Halo Center & of cells &  &  & Pressure & Met. & RV & Cont. & Cont. \\
 & & [Msol] & [kcp3] & [kpc] &  & [kms] & [kpc] &  & [Zsol] & [cgs] & [cgs] & [Zsol] & [kms] & & \\
\hline
0	&	Hurricane	&	  3414.2   	&	 0.2   	&	0.3  	&	0.4   	&	-103.1               	&	19.7          	&	 61.0      	&	0.02   	&	1.9e-13  	&	1.5e-13  	&	0.02 	&	-110.9  	&	0	&	0	\\
1	&	Hurricane	&	 20119.2	&	 0.4  	&	0.5  	&	0.1   	&	-222.7               	&	19.8          	&	149.0      	&	0.02   	&	1.2e-13  	&	1.2e-13  	&	0.01 	&	-240.8  	&	0	&	0	\\
2	&	Hurricane	&	228720.8	&	 0.4	&	0.5  	&	0.3   	&	-155.5               	&	19.5          	&	155.0      	&	0.01   	&	4.1e-14  	&	3.3e-14  	&	0.02 	&	-123.1  	&	0	&	0	\\
3	&	Hurricane	&	855973.6	&	 1.0    	&	0.6  	&	0.2   	&	-200.8               	&	18.4          	&	361.0      	&	0.01   	&	5.8e-14  	&	4.8e-14  	&	0.04 	&	-165.6  	&	0	&	0	\\
4	&	Hurricane	&	 89661.0 	&	 0.06	&	0.2  	&	0.1   	&	-270.4               	&	14.6          	&	 23.0      	&	0.01   	&	8.6e-14  	&	6.4e-14  	&	0.01 	&	-267.8  	&	0	&	0	\\
...    	&	 ...  	&	         ...     	&	      ...    	&	     ...  	&	         ...   	&	       ...               	&	      ...          	&	  ...      	&	    ...   	&	         ...  	&	         ...  	&	     ... 	&	        ...  	&	...  	&	...  	\\
25954	&	Tempest	&	   655.6    	&	 4.0	&	1.0  	&	0.0   	&	 246.5               	&	70.8          	&	 24.0      	&	0.6   	&	7.3e-16  	&	7.0e-16  	&	0.7 	&	 258.7  	&	0	&	0	\\
25955	&	Tempest	&	  3650.9     	&	21.8    	&	1.7  	&	0.1   	&	  94.4               	&	68.7          	&	132.0      	&	0.2   	&	6.6e-16  	&	6.1e-16  	&	0.3 	&	  95.7  	&	0	&	0	\\
25956	&	Tempest	&	   762.9    	&	 4.6    	&	1.0  	&	0.3   	&	  63.6               	&	75.4          	&	 28.0      	&	0.2   	&	1.1e-15  	&	8.2e-16  	&	0.4 	&	  75.5  	&	0	&	0	\\
25957	&	Tempest	&	  2114.9     	&	25.4    	&	1.8  	&	0.3   	&	 352.1               	&	78.0          	&	154.0      	&	0.3   	&	1.5e-16  	&	1.1e-16  	&	0.6 	&	 355.8  	&	0	&	0	\\
25958	&	Tempest	&	   361.2   	&	 4.3    	&	1.0  	&	0.2   	&	 305.2               	&	79.4         	&	 26.0      	&	0.5   	&	5.0e-16  	&	4.4e-16  	&	0.6 	&	 276.0  	&	0	&	0	\\

\hline
\end{tabular}
\caption{Example table for the clump properties. The full table is available online. The columns describe, from left to right, the clump ID, the halo name, the total gas mass of the clump, the clump volume, the size as a sphericalized radius, the clump elongation, the average radial velocity, the distance of the clump center to the halo center, the number of cells in the clump, the average metallicity, the average pressure, the average pressure, metallicity and radial velocity of the clump environment, and a flag for whether the clump is a satellite/ISM or not.}
\label{tab:your_label_here}
\end{sidewaystable}

This section examines the sizes, shapes, masses, ionization states, metallicities, and other clump properties. We begin with bulk properties and population statistics for the ensemble of clumps from all six FOGGIE halos. Figure \ref{fig:clump_hists} shows the distributions of radial velocity, gas mass, radius, and shape of the clump population for all six FOGGIE halos combined. We note that the selected clumps cover large ranges in nearly every property we study, suggesting that they do not have a single origin, environment, or fate. 

\subsection{Bulk Properties: Clump Sizes, Masses, Radial Velocities, and Shapes }

\begin{figure*}
    \centering
    \includegraphics[width=.99\textwidth]{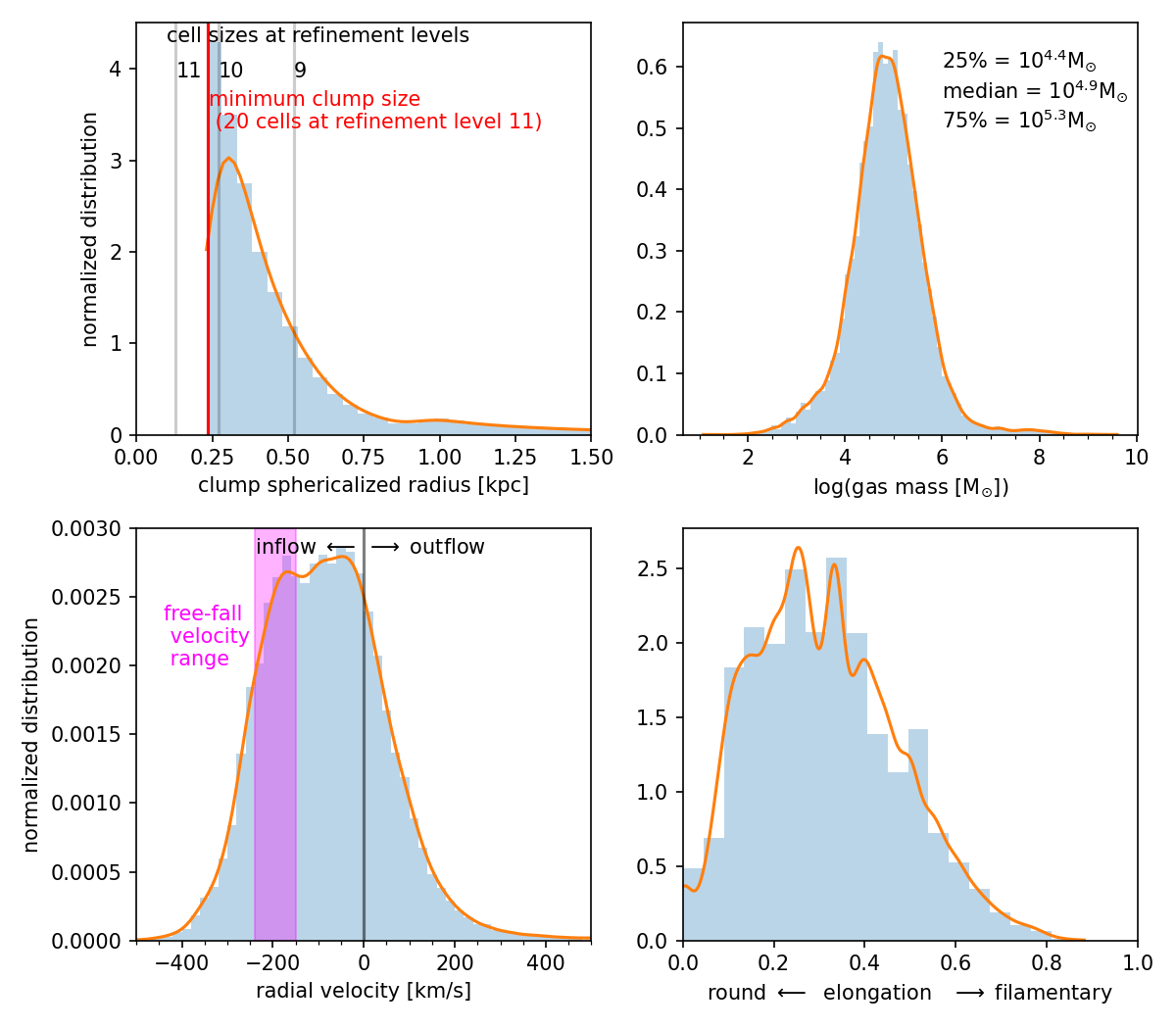}
    \caption{The radius (top-left), mass (top-right), velocity (bottom-left), and shape (bottom-right)
    of the clumps. In blue we show the histograms and in orange the kernel density estimate of the same data. For size we use a proxy, the ``sphericalized clump radius.'' Their minimum size of 0.23 kpc is defined by the minimum size we require in the clump finding process and the resolution of the simulation. The median size of the clumps is around 0.4 kpc. While the size histogram has a clear cutoff, the mass distribution follows a smooth distribution with no sign of cutoff, indicating we are not missing a significant amount of mass trapped in unresolved structures. The mass of the average clump peaks around $10^{5}~\rm{M_{\odot}}$ and are somewhat more round than filamentary, with a large scatter over possible morphologies. In terms of radial velocity we find that the majority of clumps are slowly moving towards the galaxy, with a few exceptional clumps leaving the galaxy at high velocities.}
    \label{fig:clump_hists}
\end{figure*}

{\it Sizes:} Our clump identification method allows clumps to have effectively any  shape as long as they occupy 20 or more contiguous cells. To estimate their sizes we use a ``sphericalized'' radius that approximates the clump's volume as a sphere, that is, 
\begin{equation}
    \rm{sphericalized~radius} \equiv \left(\frac{3}{4\pi} \times V_{\rm{clump}}\right)^{1/3}.
\end{equation}
The upper-left panel of Figure~\ref{fig:clump_hists} shows the cell sizes at the simulation refinement levels 9, 10 and 11 (corresponding to 548, 274, and 137 pc, respectively) in grey lines. We also show in red the minimum clump size of 230\,pc set by the sphericalized radius corresponding to a 20 cell volume at refinement level 11. The median clump size is 368\,pc.

{\it Masses:} As shown in the top-right panel of Figure~\ref{fig:clump_hists}, the median mass of a CGM clump is approximately $10^{5}$\,M$_{\odot}$. The smooth shape of the mass distribution (with no abrupt cutoffs) is a good indicator that we are not artificially omitting lower-mass clumps by enforcing the 20-cell minimum volume in the final step of selection. 
{We note, however, that the peak of this distribution may well be resolution-dependent.}
The second peak just below $10^{8}$\,M$_{\odot}$ is due to star-forming clumps in the ISM of the central galaxies, as we show in Figure~\ref{fig:hexplot2} and discuss in \S\ref{sec:ismcontam}.

{\it Radial Velocities:} We find that clumps span a wide range of radial velocities with respect to the host system's disk (lower-left panel of Figure~\ref{fig:clump_hists}). There is a skew towards negative values (infall) with a wide range extending past $v = 400$\,km\,s$^{-1}$ in both directions. Clearly density-selected clumps trace both inflowing and outflowing gas in the FOGGIE CGM, although the majority of them seem to be infalling at or just below the local free-fall velocity. We mark in pink the approximate range of typical free-fall velocities for our simulated halos.

{\it Shapes:} Elongation is an approximate indicator of the shape of the clump, whether it is more round or more filamentary. To calculate the elongation we consider the minimum ($\ell_{\rm min}$) and maximum ($\ell_{\rm max}$) extents of the structure in any of the $x$, $y$ and $z$ directions and define elongation $e$ as
\begin{equation}\label{eq:elongation}
    e = \frac{\ell_{\rm max} - \ell_{\rm min}}{\ell_{\rm max}+\ell_{\rm min}}.
\end{equation}
In this formulation more spherical clumps tend toward $e = 0$ while more filamentary clumps tend toward $e = 1$. The numerical value of elongation is dependent on the orientation of the clump with respect to the grid, but for the scope of this exploration this approximation suffices. From the resulting distribution (bottom-right panel of Figure~\ref{fig:clump_hists}), we infer that there is a large distribution in shapes but that the average clump is more round than filamentary. The distribution peaks at $e\sim 1/3$, which would correspond to a clump that is twice as long as it is wide in the other two directions.

\subsection{Physical properties of the clumps and their immediate environments}

\begin{figure*}
    \centering
    \includegraphics[width=\textwidth]{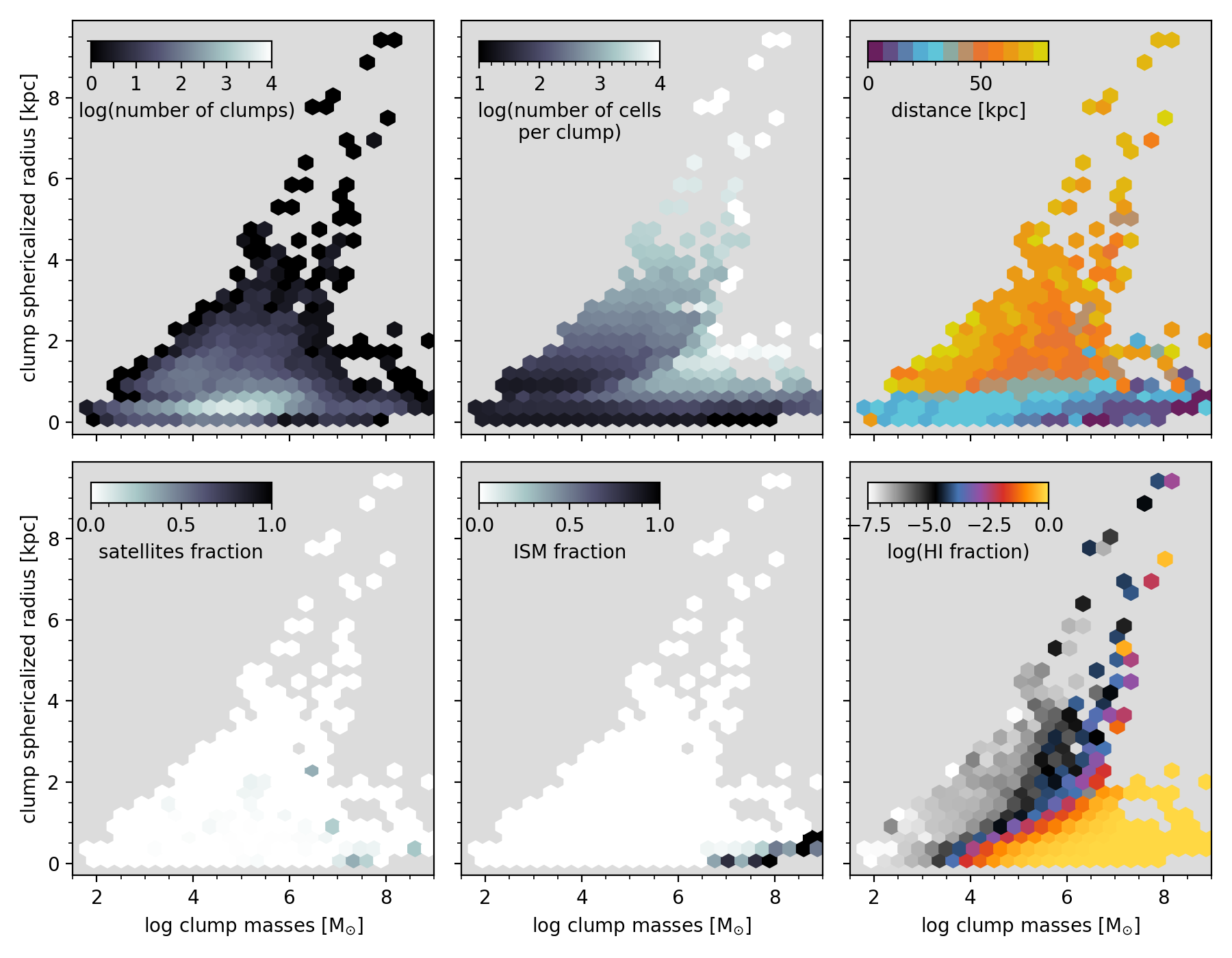}
    \caption{These six panels show how various properties of the selected clumps vary with the clump size (using the sphericalized radius as a proxy) and with clump mass. The figure shows clumps from all six halos aggregated together. Each 2-D distribution is color coded separately. From top left to bottom right, we show: the number in each 2-D bin, the number of gas cells per clump, their distance to the galaxy center, the fraction of satellites in the clump population, the fraction of ISM clumps in the total clump population and the average \hi\ fraction in each size-mass bin. 
    Further details are described in the text in Section~\ref{sec:global}.}
    \label{fig:hexplot1}
\end{figure*}

We now explore the internal physical variables inside clumps and compare them to the nearby CGM environment. Since the clumps are selected purely on the density field they provide otherwise unconstrained probes of the other physical variables (temperature, metallicity, etc.).
We investigate how the dense clumps of the CGM trace these other fields. To visualize these results we explore various clump properties in the mass-size plane in Figures \ref{fig:hexplot1} and \ref{fig:hexplot2}.
We start in Figure~\ref{fig:hexplot1} by showing the total number of clumps and the number of cells-per-clump (i.e., a measure of how well-resolved the clumps are) in this parameter space to guide future discussions. Because our selection criterion, by design, did not make an a priori assumption about whether a clump was in ``the circumgalactic medium'' or not, we also show in Figure~\ref{fig:hexplot1} which regions in this plane are potentially contaminated by contributions from satellite galaxies or the ISM. Figure~\ref{fig:hexplot2} then explores the clumps' radial velocities, pressures, and metallicities, and how these quantities compare to that of their immediate environs.
Throughout, we define the environment as a spherical region around the clump, with the diameter being the maximum extent of the clump in any spatial direction plus two additional cells. From this sphere we subtract the clump itself and treat the remainder as its environment.

\begin{figure}
    \centering
    \includegraphics[width=0.48\textwidth]{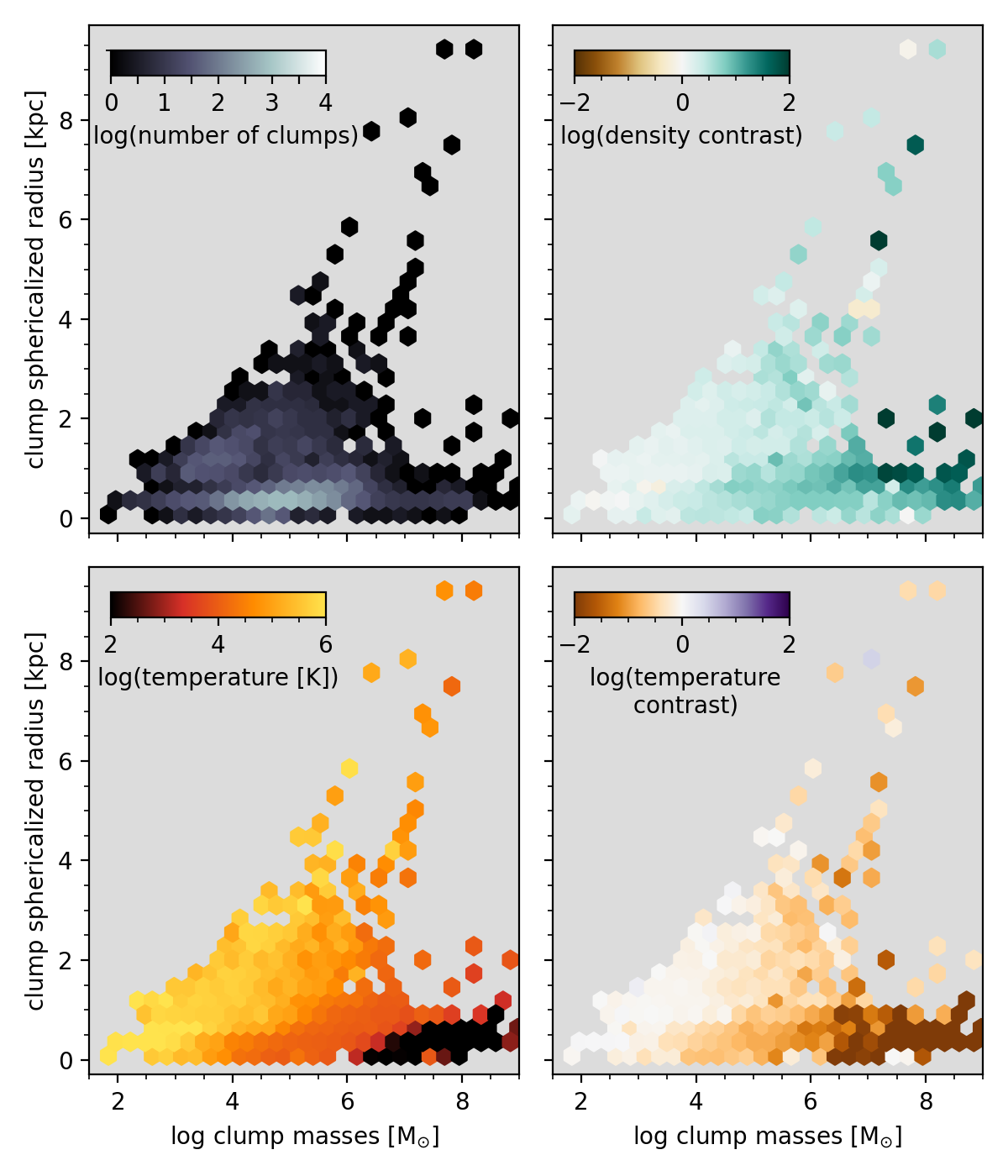}
    \caption{These four panels show how density and temperature of a subset of the clumps vary with the clump size (proxied by the sphericalized radius) and with clump mass. The figure shows clumps from all six halos aggregated together. Each 2-D distribution is color coded separately. 
    The top left panel shows the number of clumps in each hexbin for this subset.
    The top left panel shows the density contrast of the clumps to their ambient medium. 
    The bottom left panel shows the average temperature in each clump bin and the bottom right panel shows the temperature contrast.  
    Further details are described in the text in Sections~\ref{sec:density} and \ref{sec:temperature}. }
    \label{fig:hexplot3}
\end{figure}

\begin{figure*}
    \centering
    \includegraphics[width=\textwidth]{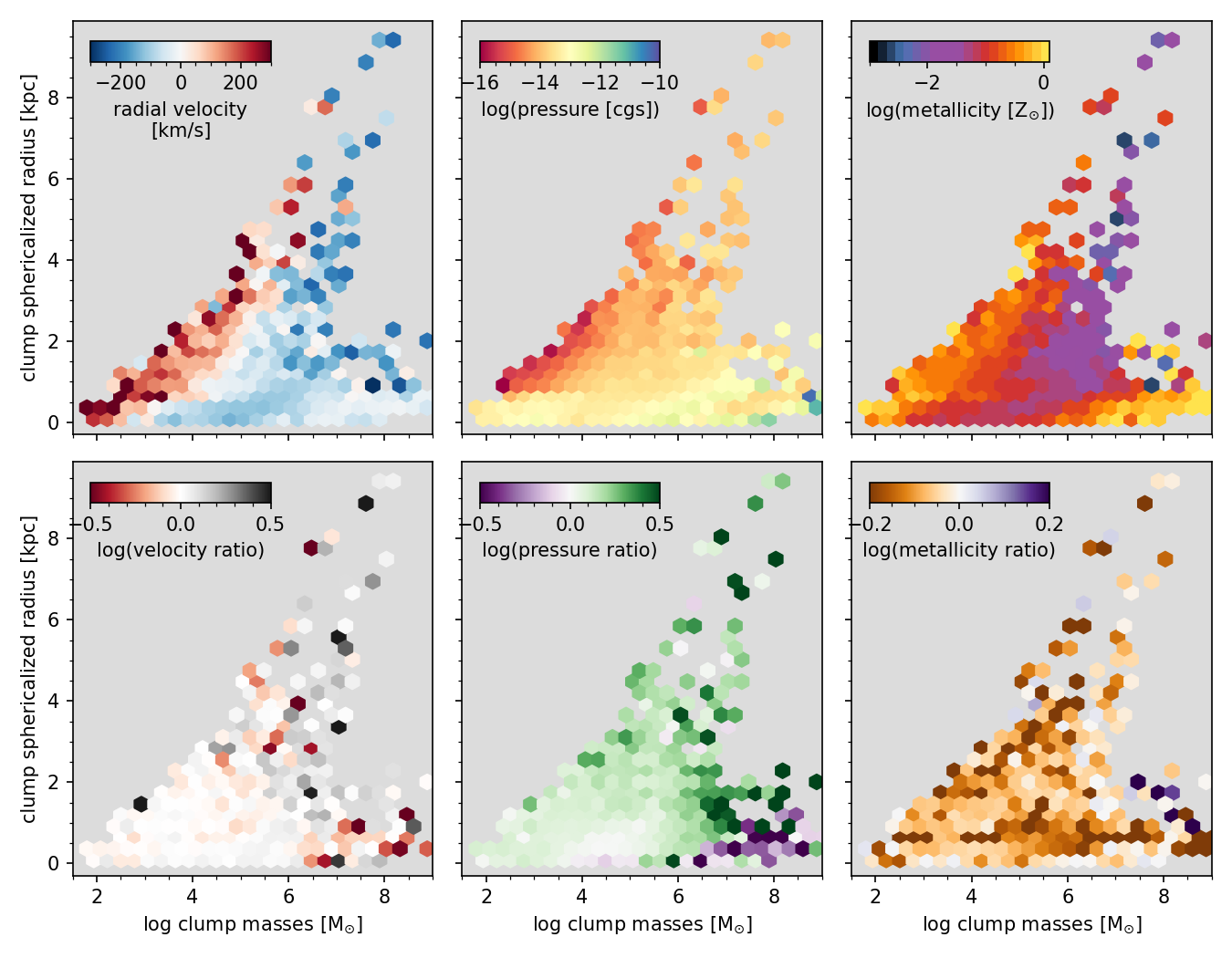}
    \caption{These 6 panels show how various properties of the selected clumps vary with the clump size (proxied by the sphericalized radius) and with clump mass. The figure shows clumps from all six halos aggregated together. Each 2-D distribution is color coded separately. The top rows show the properties of the clumps themselves, whereas the bottom panels show the same property but with respect to the clump environment. From left to right, we show: the average radial clump velocity (top) and the ratio between the average clump radial velocity and the radial velocity of the clump environment (bottom); the clump pressure (top) and the pressure ratio between clump and environment (bottom); and the clump metallicity (top) and the metallicity difference between clump and environment (bottom). The colorbars for the log ratios were chosen symmetrical around 0. To give a better idea of the total distribution, the 1/99 percentile of the pressure ratios is -0.6/0.7 with 8\% of the data outside the colorbar range. For the metallicity ratios the 1/99 percentile is -1.3/0.2 with 11\% of the data outside the colorbar range.
    Further details are described in the text in Section~\ref{sec:global}. }
    \label{fig:hexplot2}
\end{figure*}

\subsubsection{Mass-size distribution of CGM clumps}

In the top left panel of Figure \ref{fig:hexplot1} we show the same data as in Figure \ref{fig:clump_hists} but now in a 2-D plane with hexagonal bins instead of as 1-D histograms.
This allows us to appreciate the distribution of the number of clumps in the mass-size plane and helps to evaluate the other panels in Figures \ref{fig:hexplot1} and \ref{fig:hexplot2}.

\subsubsection{Resolution of simulated CGM clumps}

The top middle panel of Figure \ref{fig:hexplot1} shows the number of cells per clump in each of the hexbins, where the minimum is set by the chosen detection criterion of 20 cells. The clumps are resolved with a median of $\sim 70$ cells.
A tiny fraction ($\lesssim 0.05 \%$) have more than $10^4$ cells and are well-resolved large structures, typically found at larger radii in the CGM.

\subsubsection{Spatial distribution of clumps in the halo}

The top right panel of Figure \ref{fig:hexplot1} shows the average distance of the clumps in of each mass-size bin from the center of the halo. 
Most clumps are close to the disk, within 30--40 kpc, as can be appreciated from the halo illustrations in Figure \ref{fig:halosatz1}. In the inner halo the clumps are smaller ($\lesssim 1$ kpc) whereas in the outer halo clumpy structures can be significantly larger ($\lesssim 10$ kpc).

\subsubsection{Satellite contribution to the clump finding}

The bottom left panel of Figure \ref{fig:hexplot1} shows the fraction of 
clumps in each hexbin that are cospatial with satellites, i.e., which are likely not ``true'' circumgalactic clumps.
We define satellites by the presence of a subhalo, as identified by ROCKSTAR halo finder \citep{Behroozi2013}. For details on how we identify satellites and substructures in the FOGGIE simulations, see \citet{Wright2024}. 
{We flag any clump belonging to a satellite. We find that the gas associated with satellites is typically confined within 10\%\ of the subhalo virial radius, which is the cutoff we use for clump identification with a satellite.}
From Figure~\ref{fig:hexplot1} we see that the vast majority of structures we identify in these halos are CGM structures, with little contamination from satellites except at $\gtrsim 10^{7}$\,\Msun.

\subsubsection{ISM contribution to the clump finding}\label{sec:ismcontam}

Similarly, the bottom middle panel of Figure \ref{fig:hexplot1} shows the fraction of ISM contaminants in each hexbin. 
We define here ISM as any gas cell that contains young stars (stellar ages less than 8 million years), which typically spans the central few kpc of each galaxy. 
If any of our clumps fall within this region, we flag it as ISM contamination.
We notice that all the ISM structures are clearly clustered at the high mass ($\gtrsim 10^{6.5}$\,\Msun) and small size ($\lesssim 0.5$\,kpc) region of this parameter space. Therefore, any trend deviating from the majority of CGM clumps in this region can be attributed to the ISM. For example, the top-left panel of Figure~\ref{fig:hexplot1} shows that the region of the size-mass parameter space most contaminated by ISM gas has clumps that are significantly closer to the galactic center than the majority of the clumps. Likewise, as we discuss below, these clumps are more metal-rich and over-pressured than the typical circumgalactic clump.

\subsubsection{\hi\ Fraction and Ionization}

Neutral hydrogen absorption is one of the most commonly observed tracers of CGM and IGM gas because it is detectable over a wide range of redshifts and column densities from galactic disks ($N_{\rm HI} \sim 10^{21}$ cm$^{-2}$) to the diffuse IGM ($N_{\rm HI} \sim 10^{12}$\,cm$^{-2}$). Even though we have selected clumps on {\it physical} density, this selection does not yield only clumps with high \ion{H}{1} column densities---many of our clumps meet the density criterion but are highly ionized. This diversity of ionization fractions can be seen in the lower right panel of Figure~\ref{fig:hexplot1}. 
From this distribution we observe a clear trend of \hi\ fraction with overall clump gas density (see Section~\ref{sec:density}), with the small, massive clumps having \hi\ fractions close to unity and larger and/or less massive clumps showing lower \hi\ fractions. 

We see that the vast majority of clumps have high \ion{H}{1} fractions, with a neutral fraction of $\gtrsim 10$\%. Of course, some of the structures we find are in lower density regions, but those are typically more homogeneous over larger volumes and less clumpy than the cool gas in higher density regions.  It is, therefore, expected that we find a significantly larger number of \hi-rich clumps than ionized clumps. This seems to be true at all radii and there is no clear radial trend in the \hi\ fraction. The highly ionized part in this distribution lines up with outflowing clumps (Figure \ref{fig:hexplot2}) and therefore matches expectations of highly ionized, yet clumpy, material.

\subsubsection{Density}\label{sec:density}

In Figure~\ref{fig:hexplot3} we explore  the density and temperature contrast with respect to the immediate surroundings for a subset{\footnote{We do not have the information on density and temperature contrast on the full set due to computational memory issues. So we plot here all the clumps for which this information exists in the catalogs}} of our clumps.
The top-left panel of Figure~\ref{fig:hexplot3} shows the distribution of clumps in this subset: they probe the same parameter space as the full set (Figures~\ref{fig:hexplot1}, \ref{fig:hexplot2}).
The top-right panel shows the density contrast of the clumps to their immediate surroundings.
We remind the reader that we define the environment as a spherical region around the clump itself, which has an extent that is two cells larger than the maximum extent of the clump (see \S\,\ref{sec:id}). 
In rare cases, our environment selection can include nearby dense structures unrelated to the clump, which can lead to density contrasts $<1$ (negative in log space as presented in Figure~\ref{fig:hexplot3}). This is particularly true for elongated clumps in crowded environments.
For illustration, Figure \ref{fig:individual_clump_physical2} (second row, left panel) shows an example of a clump surrounded by higher nearby dense structures, which could, depending on the environment radius, lead to a negative average density contrast.  

Overall, the density contrast increases with clump mass. 
Outflowing clumps seem to have a smaller density contrast to their immediate environment compared to inflowing clumps.
The median density contrast for all clumps is $\sim 4$, whereas the median density contrast for outflowing structures (radial velocities $> 100$\,\kms) is only $\sim 2$.

\subsubsection{Temperature}\label{sec:temperature}

In the bottom two panels of Figure~\ref{fig:hexplot3} we show the absolute average temperature of the clumps (left) and the temperature contrast to their environment (right).
We notice that the clump population is not limited to cold gas in the CGM. Rather, it spans several orders of magnitude in temperature, with the clumps close to the disk having temperatures as low as 100 K and outflowing clumps that reach average temperatures of $10^6$ K.
We also notice, that while all clumps are generally cooler than their environment, there is a trend with absolute temperature: the coldest clumps show the highest temperature contrast, and the hottest clumps show the smallest contrast in temperature.

\subsubsection{Radial Velocity}\label{sec:rv}

In the top-left panel of Figure \ref{fig:hexplot2} we investigate the average radial velocity of each hex bin, where 
we can clearly see that the highly ionized clumps and the outflowing clumps occupy the same mass-size parameter space.
We can also see that the ISM-dominated region has little radial velocity and that the majority of clumps are inflowing, which is consistent with the top left panel of Figure \ref{fig:clump_hists}.

The bottom left panel of Figure \ref{fig:hexplot2} shows the radial velocity ratio between the clump itself and its immediate surroundings. 
The ratio is defined as the average radial velocity of the clump divided by the average radial velocity of the environment.
We see that the velocity ratio is close to unity, which means that on average the clumps are comoving with their surrounding medium, even in the outflowing gas. 
The largest velocity ratios are seen in the ISM dominated regions.

Combined with the typical distance from the galaxy shown in Figure~\ref{fig:hexplot1}, we can now infer that there is a population of inflowing gas in the inner CGM that is traced by a clumpy medium rather than coherent streams. Generally speaking, the FOGGIE halos have filamentary inflow in the outer CGM that disintegrates at $\sim 50$\, kpc---right where we see a strong up-tick in the number of overdense clumps (C.\ Lochhaas, in preparation). We will use tracer fields to investigate in an upcoming study how accretion flows through the streams in the outer CGM, potentially through this clumpy inner circumgalactic medium, and onto the gaseous galactic disk (C.\ Lochhaas, in preparation).
{We discuss illustrated examples of two individual clumps in these different inflowing/outflowing structures in Sec. \ref{sec:individual}. }

\subsubsection{Pressure}
We can synthesize the differences in density and temperature to consider whether the clumps are over- or under-pressured relative to their surroundings.
Simple models of the CGM assume that the volume-filling gas exists close to hydrostatic equilibrium with the gravitational potential and accretion arises from small thermal instabilities that cool and fall in faster than they can be shredded. Previous analyses of the FOGGIE simulations show that most of the mass in the CGM is far from hydrostatic equilibrium because it is constantly stirred by accretion and feedback \citep{Lochhaas2023}. 
In all of these scenarios, pressure and the forces applied by pressure gradients are key factors in driving CGM evolution. 

The top-middle panel in Figure~\ref{fig:hexplot2} shows the pressure trends with cloud size and mass; the bottom-middle panel shows the ratio between clump and environment pressure.
The majority of clumps are generally close to gas pressure equilibrium with their immediate surroundings.
The clumps departing from this trend significantly are either dominated by the ISM ($\gtrsim 10^6$ \Msun) or are relatively large ($\gtrsim 4$ kpc). 
Over the most densely populated hexbins, a nearly constant pressure ratio is obtained with much less variation than there is in temperature or density ratios. 
This result indicates that the density-bounded clumps are not steep peaks in the pressure field. 

However, the clumps often have highly irregular shapes, along 
with immediate surroundings 
that are structured on small scales (Figure~\ref{fig:clumpfinding}). 
The presence of smaller-scale substructure at all resolvable scales leads to a loss of information when averaging  over the structure. Therefore, comparing average pressures inside and outside clumps may at times obscure a more complicated picture (see also \citealt{Lochhaas2021}). 
Nevertheless, we conclude that the mainstream clumps are in pressure balance with their local surroundings.
In the cases of absence of pressure balance, the clumps, particularly the larger and more ionized clumps tend to be overpressured and thus more likely to be expanding rather than condensing.
This raises the question how long these structures typically survive, which we further investigate in Section \ref{sec:survivability}.

\subsubsection{Metallicity}
One of the big questions in galaxy evolution is how metals are transported and mixed in the gas around galaxies. The top right panel in Figure \ref{fig:hexplot2} shows the clump metallicities in the size-mass plane. Here we see that the highly ionized, outflowing material in the upper locus of clumps is relatively metal rich, with metallicities ranging from 0.1\,Z$_{\odot}$ to 1\,Z$_{\odot}$. Additionally, high neutral fraction
clumps nearer the disk/ISM have relatively high metallicities too.
In between is a population of more metal-poor clumps ($\gtrsim 0.01$Z$_{\odot}$) with generally low neutral fractions and velocities consistent with inflow. 

It is intriguing that essentially all of the clumps are at lower metallicities than their surroundings (bottom-right panel of Figure~\ref{fig:hexplot2}). For the inflowing clumps in particular, these low metallicities (both relatively and in absolute terms) are consistent with the scenario that these clumps possibly trace the breakup of inflowing streams, as discussed in Section~\ref{sec:rv}. The metallicity difference with the surroundings could potentially be used to set a limit on the cloud lifetimes relative to the metal-mixing timescales of the CGM.

\subsection{Longevity of Circumgalactic clumps}
\label{sec:survivability}

The survival of cold clumps embedded and moving within hotter gas is an open topic of active research. Many studies focus on the disruption or survival of cold clumps embedded within a hot, fast-moving wind, as this scenario is most likely to lead to clump disruption \citep{Armillotta2017,Zhang2017,Gronke2018}. The general picture is that a hot, fast wind can shear apart a cold clump within a few cloud-crushing timescales, but radiative cooling of either the hot gas onto the clump \citep{Li2020c,Sparre2020}, or within the mixing layer between the clump and the hot wind \citep{Gronke2020,Farber2022}, can prolong the survival of the cold clump. This leads to the notion of a ``survivability criterion'' for clumps, the exact form of which is debated \citep{Kanjilal2021}. Broadly, this criterion can be defined as a balance between the timescales of destruction through shear forces and growth through radiative cooling. Because the critical physics that determines the survival or destruction of clumps occurs in the thin mixing layer between the clump and its surroundings, a high simulation resolution is required to probe the process. Most studies use a ``wind tunnel'' setup where a single, typically spherical, cloud of size $\sim 1$--100\,pc is embedded within a hot, uniform wind. In cosmological simulations like those analyzed in this work, the clumps that can be resolved are much larger, but the analytic expressions that capture the relevant timescales for destruction vs.\ mixing/cooling may still apply as the most important physical properties are likely to still be the cloud overdensity and Mach number.

\citet{Abruzzo2023} collect the various survivability criteria from other works and combine them into a single, simple expression that determines whether a clump should survive or be destroyed. To assess the survivability of the clumps in the FOGGIE simulations, we use their criterion, which shows that clumps are expected to survive when
\begin{equation}\label{eqn:survive}
\alpha t_{\rm shear} \gtrsim t_{\rm cool},
\end{equation}
where $\alpha$ is a non-uniform constant on the order of a few to $\sim 10$.
Physically, this criterion represents the concept that clumps will grow (and thus not be destroyed) when the cooling time is short enough for gas in the mixing layer to be able to be added to the cloud instead of advecting past it---while removing some of the clump's original material.
The shearing time $t_{\rm shear}$ is defined as the clump size divided by the velocity difference between the clump and the ``wind,'' which we take to be the velocity difference between the clump and its environs (note that this is {\em not} radial velocity as shown in Figure~\ref{fig:hexplot2}, but rather the full velocity field).
We calculate the velocity difference $\Delta v$ from the average velocities $\overline{v}$ in the clump and its environment by accounting for each of the three cardinal directions, with
\begin{equation}
    \Delta v \equiv \sqrt{\sum_{i}\left(\overline{v}^{\rm{clump}}_i - \overline{v}^{\rm{environment}}_i\right)^{2}} \ {\rm{for}} \ i = x, y, z.
\end{equation}
Note that this survivability criterion was derived assuming supersonic wind speeds (in which the clump is initially at rest) and clump-background density contrasts of $\chi\gtrsim100$, ratios we find for $\ll 0.1$\% of our clumps.  This criterion is thus a rough estimate for our clumps, and the value of $\alpha$ captures deviations from these assumptions. Some high-resolution simulations of the mixing layer instead use the turbulent eddy turnover timescale  of $t_{\rm shear}$ \citep{Fielding2020a,Tan2021}. However, we choose to use $t_{\rm shear}$ for the clumps, as we do not resolve the mixing layer.
The relevant cooling time is the minimum cooling time in the mixing layer, i.e., at the edge of the cloud. We therefore define here a clump's $t_{\rm cool}$ to be the minimum cooling time in the cells immediately surrounding the clump. 
This choice is to make sure that we do not consider cooling times within the clump itself but only within the mixing layer surrounding the clump.

\begin{figure}
    \centering
    \includegraphics[width=.5\textwidth]{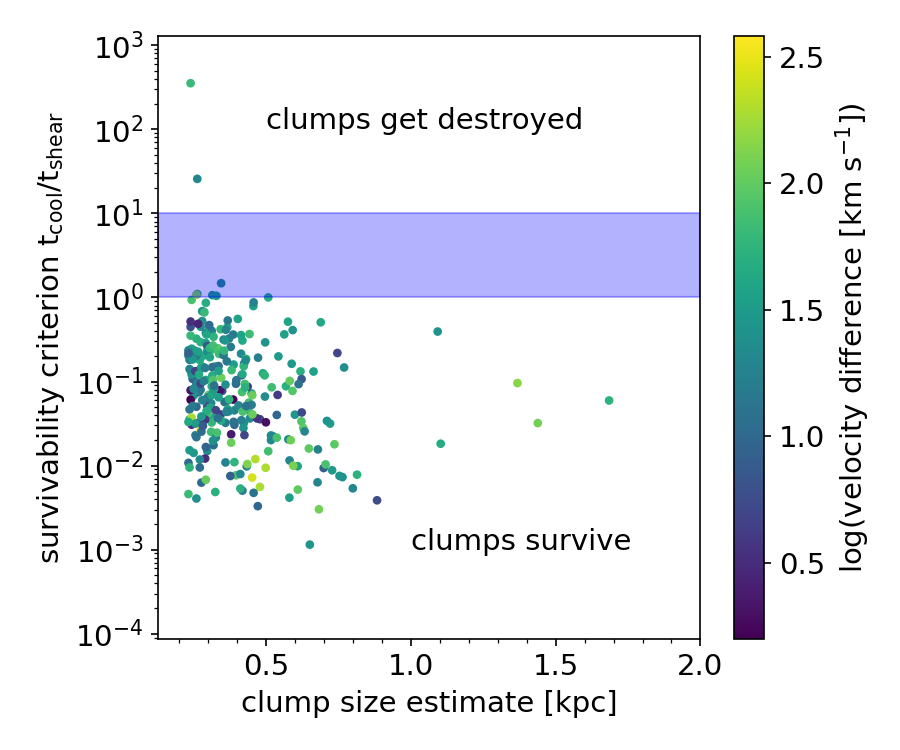}
    \caption{Survivability criterion $t_{\rm ool}/t_{\rm shear}$ vs clump size: We find that clumps with sphericalized radii $\gtrsim 1$\,kpc are typically long-lived, whereas smaller ones may or may not be destroyed soon. }
    \label{fig:survivability}
\end{figure}

Figure \ref{fig:survivability} shows $t_{\rm cool} / t_{\rm shear}$ versus the clump sphericalized radius for clumps in a random CGM patch of the Squall halo. A line is drawn to indicate $t_{\rm cool} = t_{\rm shear}$. We find that clumps larger than 0.5\,kpc  pass the survivability criterion, and only $\sim \, 4 \%$ of the clumps $\lesssim 0.5$\,kpc are expected to be destroyed within a few cooling times. We stress that the survivability criterion is not a hard line at $t_{\rm cool} = t_{\rm shear}$, but may vary based on the exact values of the density contrast, the relative speed between the clump and its surroundings, and the size of the clump.
We show this in Figure \ref{fig:survivability}, by shading the region between $t_{\rm cool} = t_{\rm shear}$ and $t_{\rm cool} = 10 \times t_{\rm shear}$.

Comparing Figure \ref{fig:survivability} to the radial velocity ratio plot (bottom-left panel) in Figure~\ref{fig:hexplot2}, we see that most of our clumps are long-lived because they have quite small velocities relative to their environs, i.e., a small $\Delta v$ leads to a large $t_{\rm shear}$. (An exception to this trend is the example clump shown in Figure~\ref{fig:individual_clump_physical1}, which has an unusually high velocity contrast with its surroundings that produces the prominent head-tail structure. Such structure is generally not seen in the other FOGGIE clumps that have slower relative velocities with their surroundings.)
The median velocity difference for the clumps in Figure \ref{fig:survivability} is only 27 \kms. 
The typical sound speed in these clumps is $\sim$15 \kms. 
In the wind-tunnel simulations that derive the survivability criterion we use here, it is generally assumed that wind speed $v_w$ is large enough to be supersonic \citep{Abruzzo2023}. The clumps in FOGGIE are not located within a hot and fast wind, but rather are moving slowly relative to their surroundings through the CGM. In addition, the density contrast $\chi$ between the clumps found in FOGGIE and their surroundings are much smaller than that typically assumed in cloud crushing simulations, with a median contrast of $\chi\sim 4$ and $<1$\% of clumps with $\chi \gtrsim 35$ instead of the typically-assumed $\chi\gtrsim100$. Assuming that the survivability criterion in equation~\ref{eqn:survive} still holds under typical FOGGIE conditions, it is not surprising that most FOGGIE clumps are expected to survive.
{Of course, given that the mixing layer itself is not resolved in FOGGIE it remains unclear if the derived life times are potentially overestimated, since the prescription for survival was derived from clouds where the turbulent mixing layer was resolved, capturing the small-scale physics happening at the interface between the clouds and their ambient medium.
Generally, we expect clumps that move against their environment, such as illustrated in Fig. \ref{fig:individual_clump_physical1} to have shorter lifetimes compared to those that are well embedded within their larger envelope (Fig. \ref{fig:individual_clump_physical2}). }

In future work, we will use tracers to follow the evolution of identified clumps from one snapshot to the next to empirically gauge their survival. 
Clumps in cosmological CGM environments tend to be larger, move more slowly, and be less dense than in wind tunnel simulations.
By investigating the survival of cosmological clumps, 
we will be able to determine the validity of the above survival criterion as applied to a cosmologically-simulated CGM. 

Meanwhile, \citet{Ramesh2024OE} have made use of the tracking capabilities in the GIBLE simulations to determine the origin and evolution of cold gas cloud structures and find $\lesssim$ 10\%\ of CGM clumps in their simulations are long-lived, with the large majority undergoing frequent shattering and merging processes. 
We note here that their selection for clumps is different from ours (where we select on local overdensities, they select on a temperature cut, see Section \ref{sec:comparesims} for more details) which could result in a different cloud population and makes direct comparison somewhat challenging.
With the tracking capabilities planned for future FOGGIE runs, a more robust comparison between those different populations can be made.

\subsection{Clump Mass Function}
\label{sec:CMF}

\begin{figure}
    \centering
    \includegraphics[width=.5\textwidth]{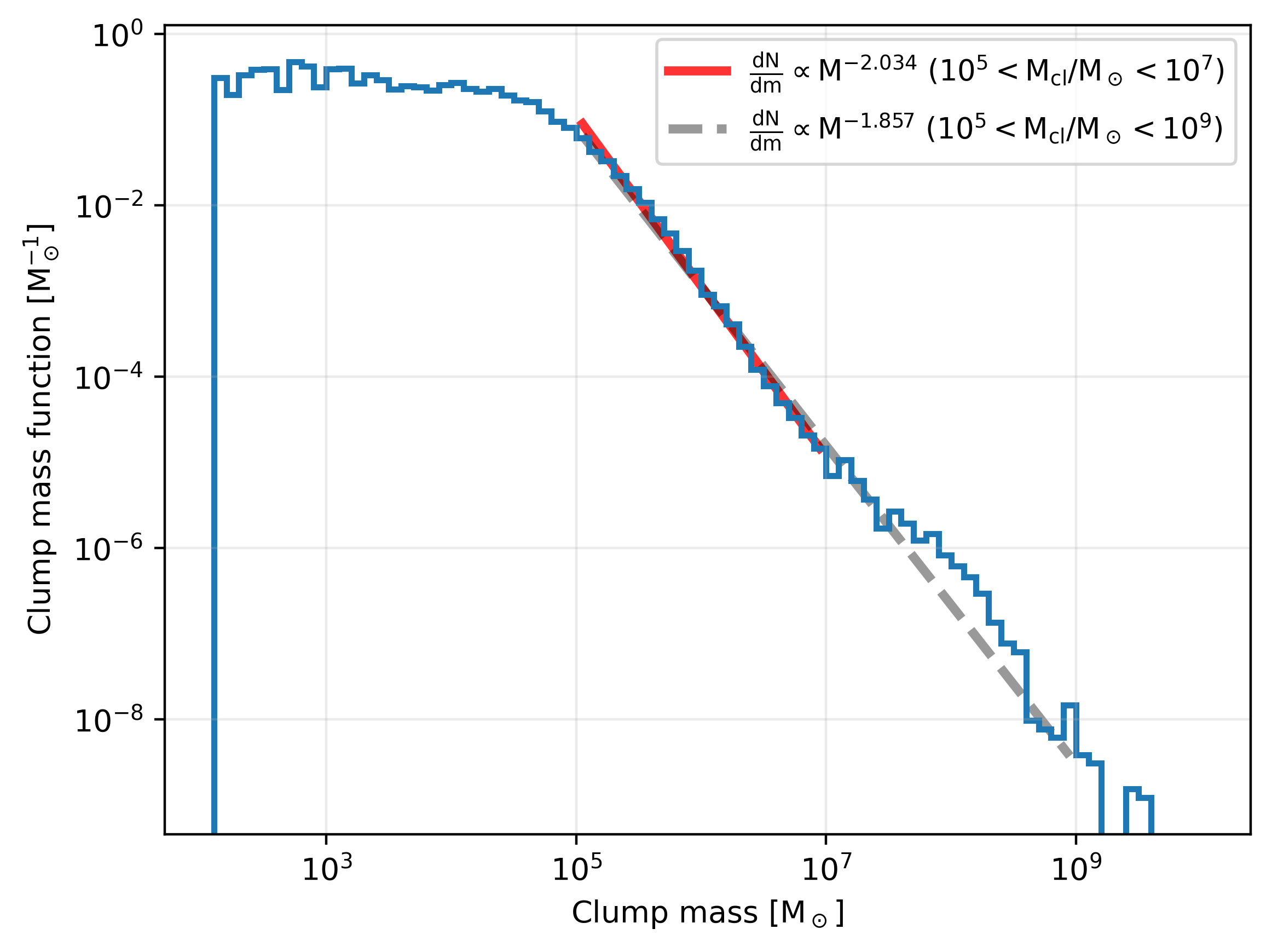}
    \caption{Clump probability density distribution ($\mathrm{dN/dm}$) as a function of cloud mass for all clumps in our sample.  The red line is a weighted fit to the distribution for clump masses in the range $\mathrm{M_{cl} = 10^5 - 10^7~M_\odot}$, which has a slope of $\mathrm{dN/dm \propto M^{-2.034}}$ and is normalized to the data at $\mathrm{M_{cl} = 10^6~M_\odot}$.  The grey dashed line is a weighted fit to the distribution for clump masses in the range $\mathrm{M_{cl} = 10^5 - 10^{9}~M_\odot}$, which has a slope of $\mathrm{dN/dm \propto M^{-1.857}}$ and is normalized to the data at $\mathrm{M_{cl} = 10^6~M_\odot}$.}
    \label{fig:CMF}
\end{figure}

Figure \ref{fig:CMF} shows the clump probability density distribution function ($\mathrm{dN/dm}$) as a function of cloud mass for all the clumps in our sample.  We perform a weighted fit (weighting by $1/\sigma(m)$ for each bin) to a power law for the massive end of this distribution over the mass range $\mathrm{M_{cl} = 10^5 - 10^7~M_\odot}$ (with $N=10{,}483$ total clumps in that mass range) and power law index of $\alpha = -2.034$.  We chose this mass range as our fiducial fitting range because it excludes lower-mass clumps whose completeness is likely impacted by resolution effects, and the few number of high mass clumps ($N=277$ above $10^7$~M$_\odot$) where small number statistics are likely to impact the fit.  To test this hypothesis, we perform an additional weighted fit over the mass range $\mathrm{M_{cl} = 10^5 - 10^9~M_\odot}$ ($N=10{,}756$ clumps) and find a slightly shallower power law with an index of $\alpha = -1.857$.
Both of these power law indices are consistent with the cloud mass function of approximately $\mathrm{dn/dM \propto M^{-2}}$ that was found in the high-resolution cosmological simulation of \citet{Ramesh2024}\footnote{Note that \citet{Ramesh2024} plot N(m) rather than the more standard $\mathrm{dN/dm}$, therefore the power-law index in their Figure 9 is $\alpha \simeq -1$ rather than $\alpha \simeq -2$; if plotted as $\mathrm{dN/dm}$ the power-law index would be $\alpha \simeq -2$.}.  It is also consistent with cloud structures found in more highly idealized simulations of galactic outflows \citep[e.g.,][]{Warren2024}, ``wind tunnel'' simulations of clouds being destroyed in a supersonic wind \citep[e.g.,][]{Gronke2022}, and both hydrodynamic and magnetohydrodynamic turbulent box simulations \citep[e.g.,][]{Fielding2023,TanFielding2024}.

\section{How Individual Clumps Reflect the Properties of the Circumgalactic Medium}
\label{sec:individual_clumps}

We now consider how the clumps identified in Section~\ref{sec:id} and characterized in Section~\ref{sec:global} contribute to the full physical and observable properties of the circumgalactic medium. First, in Section~\ref{sec:individual} we consider in detail the properties of two different, but illustrative, clumps. Then, in Section~\ref{sec:fracs} we look at how the clumps as a population contribute to a halo's gas mass, volume, metal mass, and \hi\ mass.

\subsection{Dissecting Individual Clumps}
\label{sec:individual}

\begin{figure*}
    \centering
    \includegraphics[width=.99\textwidth]{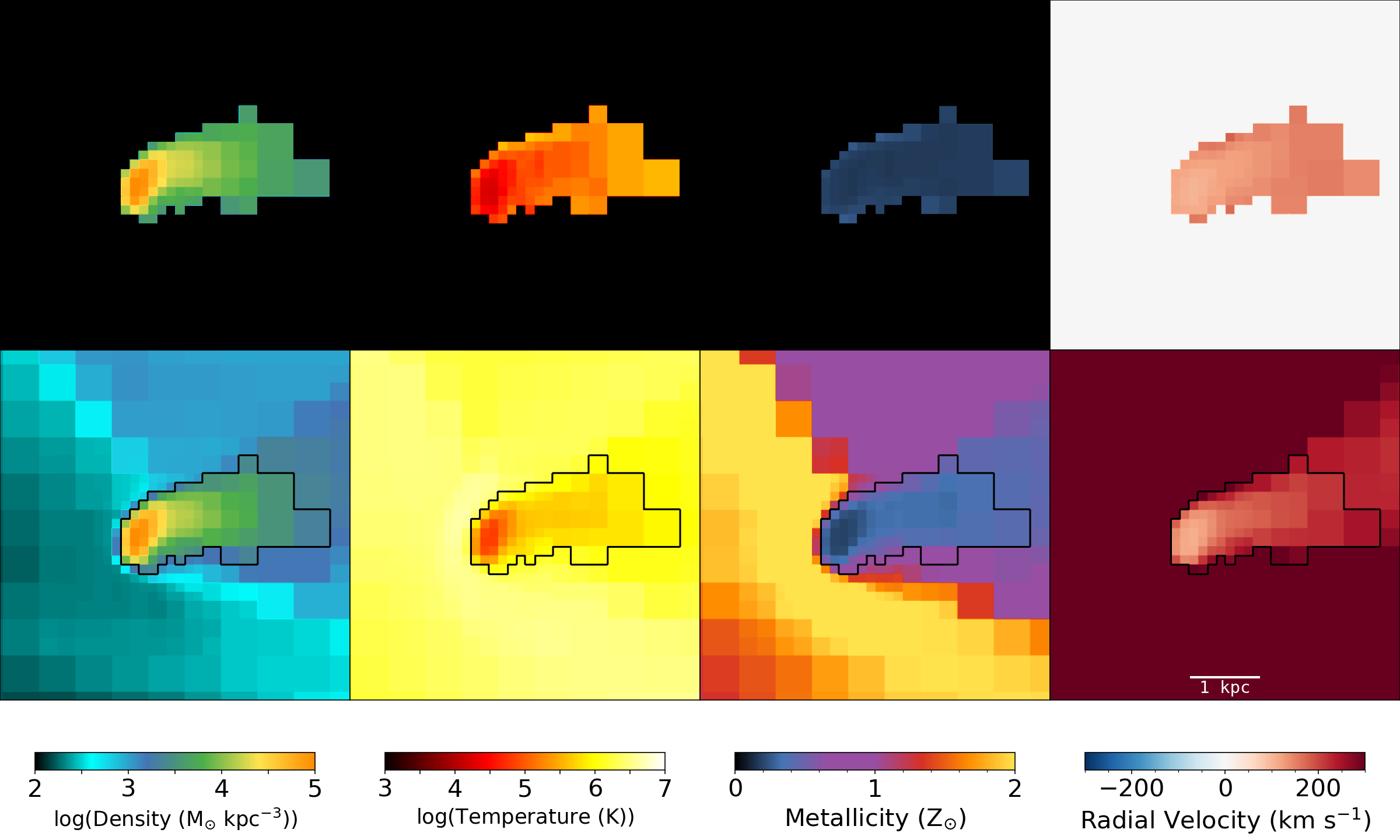}
    \includegraphics[width=.99\textwidth]{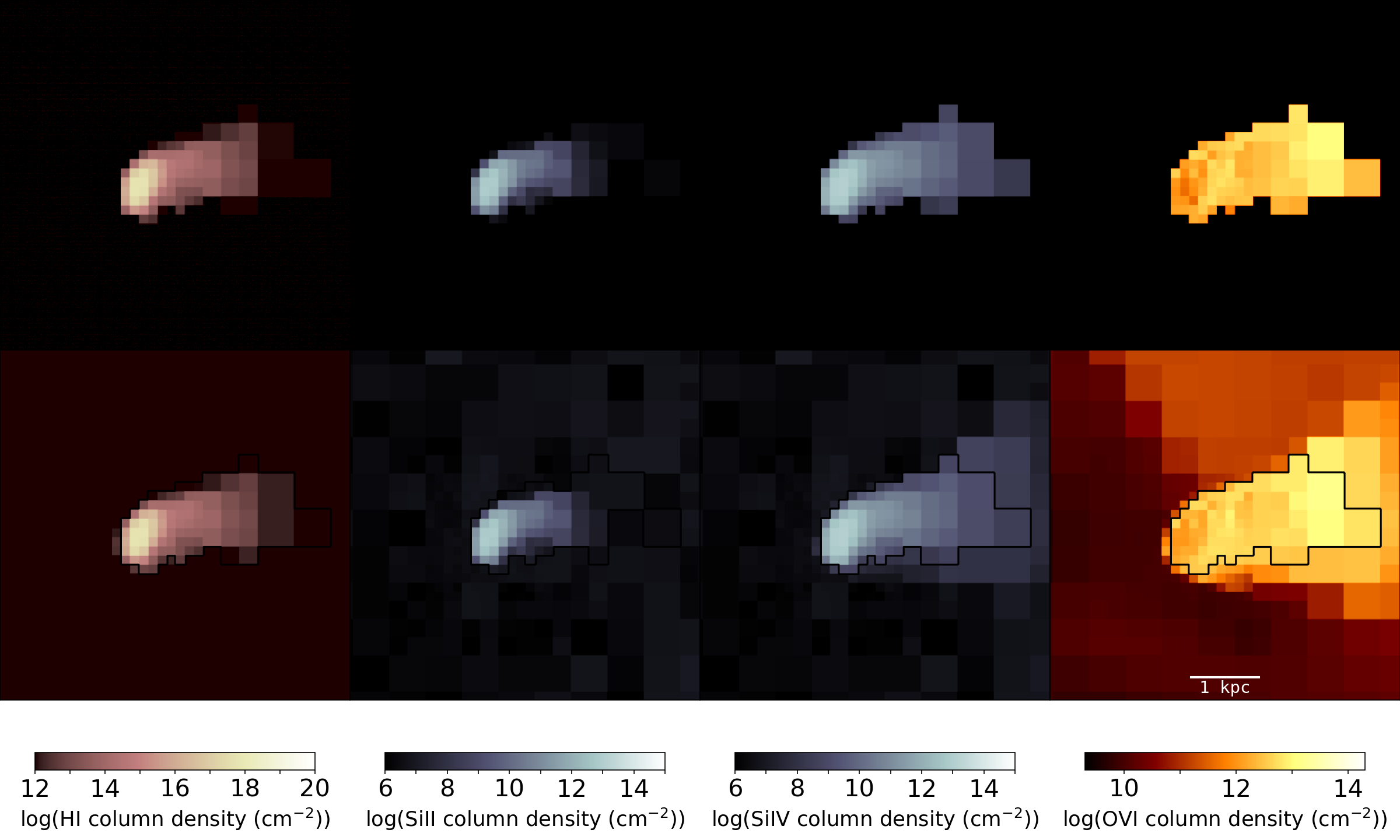}
    \caption{Physical properties of an individual clump entrained in an outflow. We find that the overdense structures show up prominently in the temperature and metallicity plots as a low temperature - low metallicity objects. In the radial velocity plot we find that while both the clump and its surroundings are outflowing, the clump itself is moving against its environment, causing the distinct head-tail structure. The \ion{H}{1} is well contained within the clump whereas high ions like \ion{O}{6} is more spatially extended and most prominent at the edge of the clump.}
    \label{fig:individual_clump_physical1}
\end{figure*}
\begin{figure*}
    \centering
    \includegraphics[width=.99\textwidth]{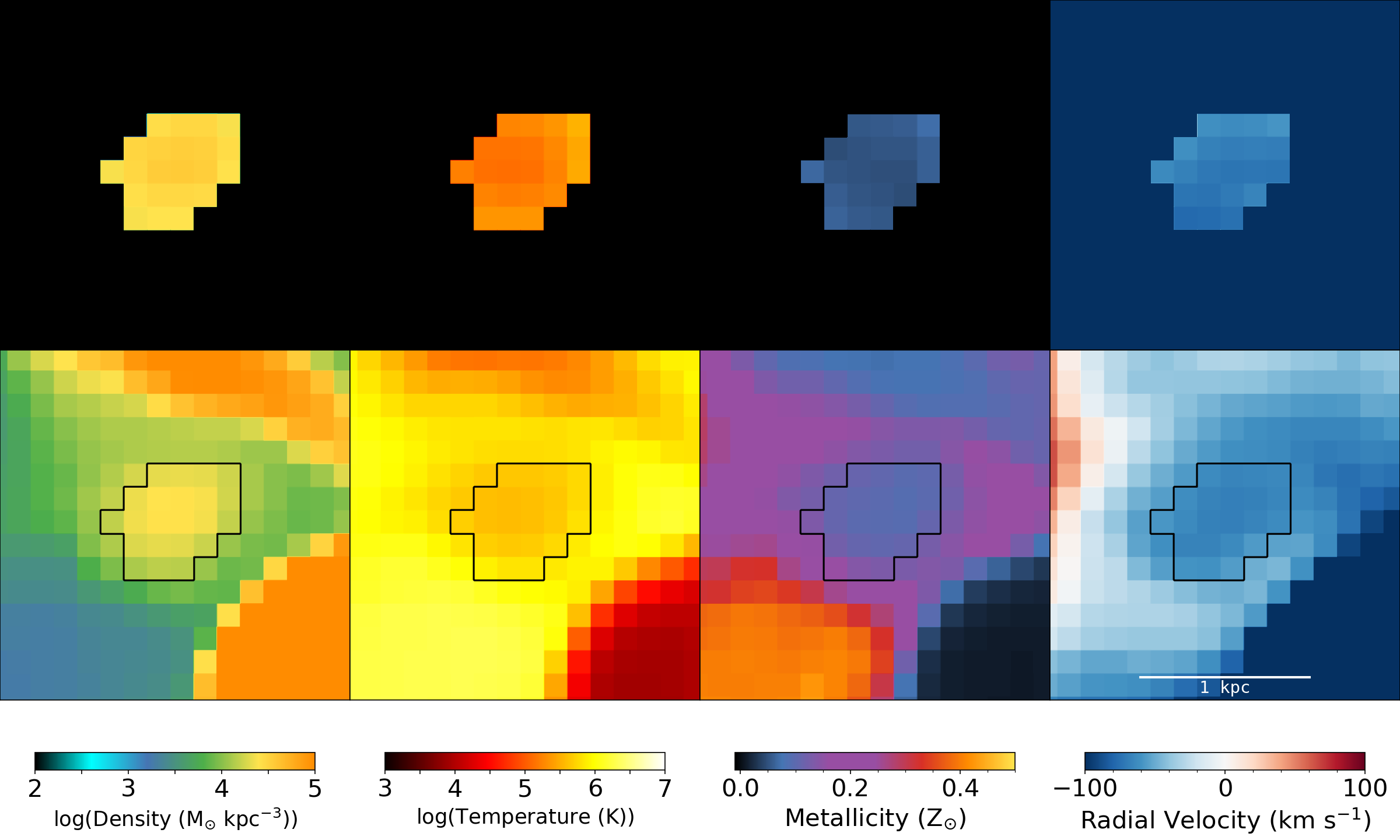}
    \includegraphics[width=.99\textwidth]{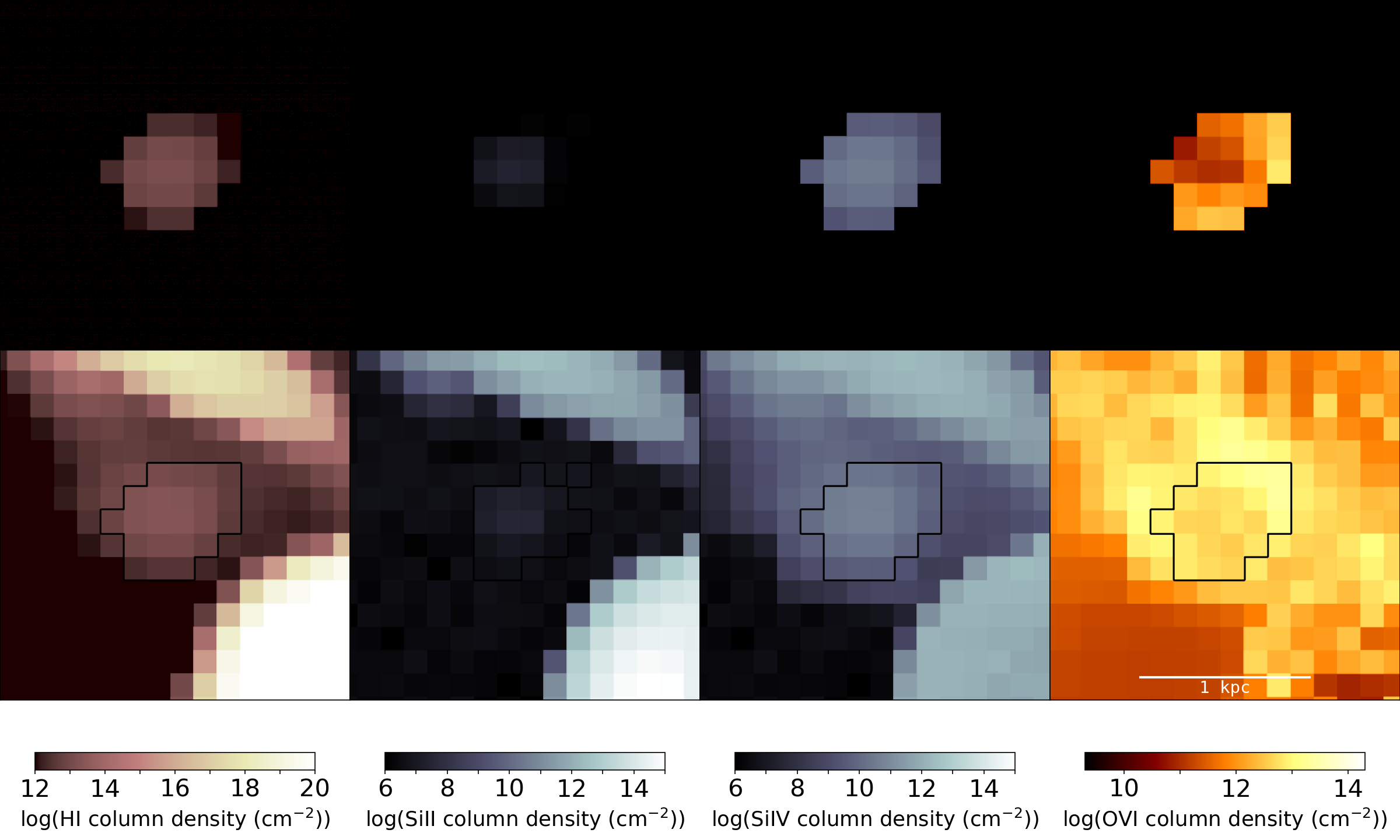}
    \caption{Physical properties of an individual inflowing clump. Similar to the outflowing case (Figure~\ref{fig:individual_clump_physical1}), we find that the overdense structures show up prominently in the temperature and metallicity plots as a low temperature, low metallicity object. In the radial velocity plot we find little distinction between the clump and its environment, suggesting it is well embedded within its surroundings and not experiencing a lot of shear. The \ion{H}{1} is well contained within the clump whereas high ions like \ion{O}{6} show up homogeneously in the projection, as there is little added \ion{O}{6} from the clump structure. In fact, the \ion{O}{6} column densities increase outside the edges, indicating that high ions trace the diffuse medium and edges of overdense structures rather than the clumps themselves.}
    \label{fig:individual_clump_physical2}
\end{figure*}

Here we consider in detail two example clumps with an eye toward their internal structure, relation to their environs, and ionization structure.
For this we extract small cubic boxes (scale bar for size estimate in the panels) from the CGM that are centered on the clump and investigate a number of physical and observable properties of the clump itself and in combination with its environment.
{To calculate the ion number densities and column densities we post-process our simulated snapshot with \textsc{Trident} \citep{TridentRef}, which uses \textsc{Cloudy} \citep{Ferland2017} to generate ion abundances for each gas cell in the simulation assuming collisional and photoionizational equilibrium. We assumed a \citet{HM2012} UV background for these calculations.}

Figure \ref{fig:individual_clump_physical1} the first example, a relatively massive (gas mass $= 10^{4.6}~\rm{M_{\odot}}$) clump with its centroid 34 kpc from the center of Tempest and a sphericalized radius of 1.4 kpc. The top panels show physical properties (density, temperature, metallicity, and radial velocity) and the bottom panels show ionic column density (\hi, \siii, \siiv, and \ovi). In both sets of panels, the top row shows the clump in isolation while the bottom row includes the clump's environment. Although this clump is not prototypical, we highlight it as it shows a classic ``head-tail'' structure seen in some Milky Way high-velocity clouds \citep[e.g.,][]{Bruens2000} and is in a relatively low-density environment. There is a clear well-resolved internal structure, with denser, colder, lower-ionization gas in the ``head'' and warmer, lower density gas more traced by \siiv\ and \ovi\ in the ``tail.'' In general there is a layered ionization structure, with the clump appearing larger as the ionization state increases.

Figure \ref{fig:individual_clump_physical2} shows a second example: a typical low mass (gas mass $=  10^{3.4}~\rm{M_{\odot}}$)  inflowing clump, 13 kpc from the center of Tempest with a sphericalized radius of 0.5 kpc. This clump is in a more crowded environment than the one highlighted in Figure \ref{fig:individual_clump_physical1}. 
Here, the clump is slightly cooler and slightly metal-poor relative to its {\em immediate} environment, as is typical for most of our clumps (Figures~ \ref{fig:hexplot3}, \ref{fig:hexplot2}). 
Also typical is the clump having approximately the same radial velocity as its immediate surroundings. However, there are prominent dense, cold, and low-metallicity structures nearby. The lower-right structure of which is likely a cold inflowing stream, as indicated by its high radial velocity. As with the previous example, the clump is larger in \ovi\ than in \siiv, and larger in \siiv\ than in \siii. Because of the crowded environment and the clump's slightly-lower metallicity, this clump would be unlikely to be visually detected in a column density projection map, however. Combined with its low velocity difference, it is doubtful that this clump would be identified as an individual ``absorber'' in a line-of-sight pencil-beam spectrum.

\subsection{Fractional occupancy of clumpy/structured vs diffuse/unstructured gas in the halo}
\label{sec:fracs}

\begin{figure}
    \centering
    \includegraphics[width=.5\textwidth]{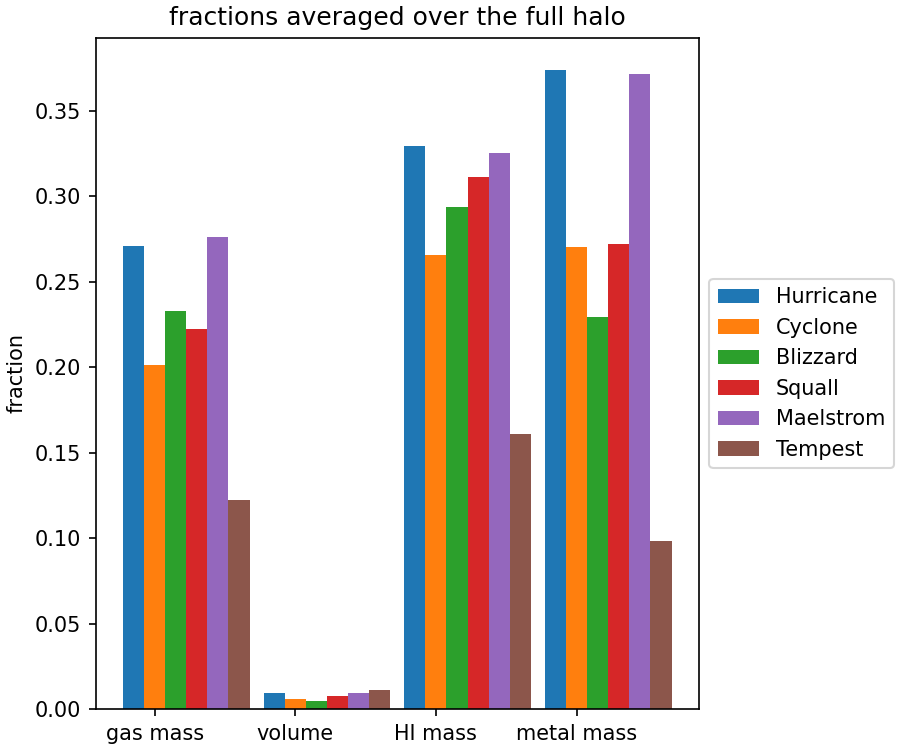}
    \caption{Fractions of gas mass, volume, \hi\ mass, and metal mass in clumps for each halo.}
    \label{fig:fractions_fullhalo}
\end{figure}

\begin{figure}
    \centering
    \includegraphics[width=.5\textwidth]{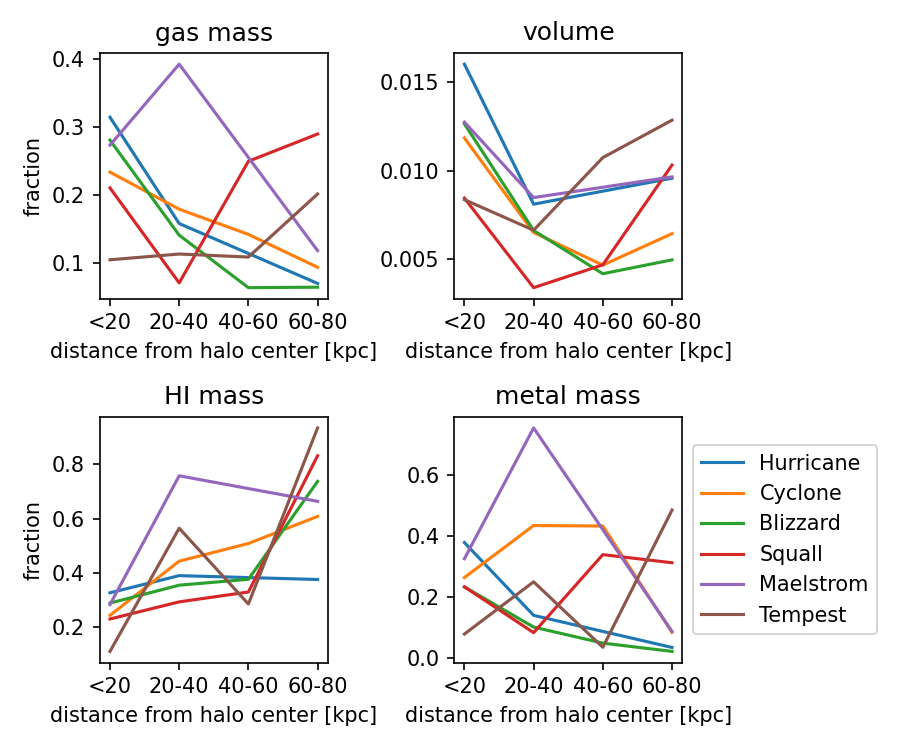}
    \caption{Fractions of gas mass, volume, \hi\ mass and metal mass in clumps for each halo split by radial distance from the central galaxy.}
    \label{fig:fractions_byradius}
\end{figure}

The projections in Figure \ref{fig:halosatz1} suggest that clumpy structure fills out much of the gaseous halos. With the clump catalogs, we can quantify how much 
of the mass and volume is occupied by
our identified, overdense clumps. Figures \ref{fig:fractions_fullhalo} shows the fractional mass, volume, \hi\ mass, and metal mass (with color-coding identical to Figure \ref{fig:denscuts}), for the full volume of each halo that was searched for clumps (out to a galactocentric radius of 80\,kpc). Across all six halos we find that on average $\sim 20$\% of the CGM mass and only $\sim 1$\% of the volume lie in identified clumps. The clumps preferentially trace gas with slightly lower ionization than the full CGM, so that the \hi\ mass fractions rise to $\sim 30$\%. The metal mass fractions are similar to the total mass fractions, with some halos exhibiting a mass-excess of metals (e.g., Hurricane and Cyclone) and others a small deficit (e.g., Blizzard and Tempest). These comparisons suggest that the clump populations of the six FOGGIE halos may have diverse physical origins, or at least different balances of inflow/outflow, enriched vs.\ metal-poor, or lower vs.\ higher ionization. This interpretation is supported by the lack of trends in the mass and volume fractions plotted in Figure~\ref{fig:fractions_byradius}. While there is a general trend toward more clumps at lower radii the volume and mass fractions occupied by the clumps do not change monotonically with radius.  This behavior probably reflects the complex large-scale structures---filaments, stripped satellite gas, and/or highly extended disks---that break what might otherwise be a spherical symmetry to the clump distribution. From Figure \ref{fig:halosatz1} it is evident that small clumps preferentially form in and around these larger structures, or that these larger structures create and/or host a fraction of the small clumps. 
This result suggests that observations tracing small-scale clumps might also trace larger, more coherent structures as well.

\begin{figure}
    \centering
    \includegraphics[width=0.48\textwidth]{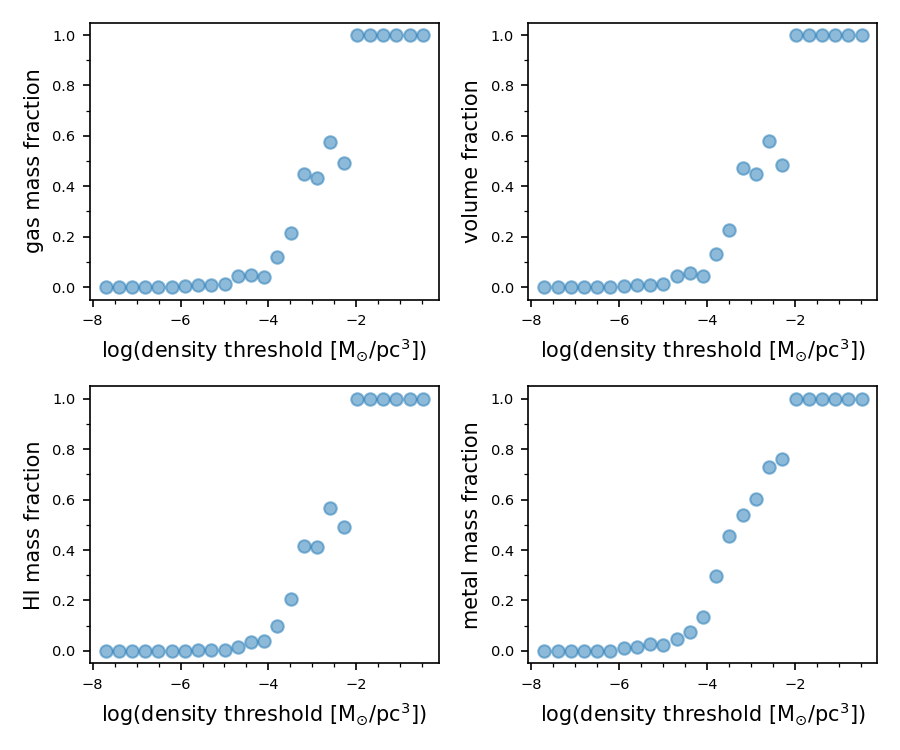}
    \caption{Gas mass, volume, \hi\ mass, and metal mass as a function of threshold density for Tempest in the range from 20--40 kpc. This radius bin is where most of the clumps are found, and has no contamination from ISM. These curves show that most of the clump mass and volume lies in the range $10^{-4}$ to $10^{-2}$ \Msun\,pc$^{-3}$, which is directly related to the density ranges in Figure \ref{fig:denscuts}. Below this range the gas is mostly diffuse and not within clumps, whereas above this density the gas is often inside clumps but these high density regions only represents a small part of the total halo mass and volume. 
    }
    \label{fig:fractions}
\end{figure}

Figure~\ref{fig:fractions} shows how the clump fractional mass and volume depend on density. 
For this figure, we calculate the fraction of different quantities contained in the clumps at each of the density thresholds used to identify the clumps (see also Figure \ref{fig:denscuts}).
Most of the clump mass and volume fractions are gained between $10^{-4}$ to $10^{-2}$ \Msun\,pc$^{-3}$ (corresponding to 0.006 particles cm$^{-3}$ to 0.6 particles cm$^{-3}$). This is compatible, by construction, with the density ranges in Figure \ref{fig:clumpfinding}. This means that gas at lower densities ($\lesssim 10^{-4}$ \Msun\,pc$^{-3}$) is generally not found in clumps, while gas at higher densities ($\gtrsim 10^{-2}$ \Msun\,pc$^{-3}$) is mostly is found in clumps, even though those clumps occupy a relatively small fraction of the halo volume and mass (Figure \ref{fig:fractions_fullhalo}). Similar patterns are obtained in the \hi\ and metal mass fractions. 

\section{Discussion}
\label{sec:discussion}

We began with the motivation to explore small-scale structure in the CGM, guided by observational indications that such structures are widespread and may bear important physical clues to the role of the CGM in galaxy evolution. FOGGIE itself was specifically designed to reach high spatial and mass resolution in low-density gas to address such questions with an optimized numerical approach. As we will summarize below, other simulation approaches are reaching similar scales and the observations have continued to advance. The observational task is to characterize the small-scale structure in the CGM of the Milky Way and other galaxies: to constrain the mass and size distributions, to assess their metal content, kinematics, and, from these, to understand their origins. Ultimately, our goal is to use these small-scale structures as a diagnostic tool for uncovering the accretion and feedback mechanisms that couple the CGM to the star-forming regions of galaxies. With this context, we will now consider our FOGGIE results in comparison to other recent simulations and key observations.
Section \ref{sec:comparesims} compares our results to other simulations with high enough resolution to see CGM small-scale structure and Section \ref{sec:compareobs} compares our findings to high velocity clouds in the CGM of the Milky Way as well as cloud measurements from multiple sightline studies. 

\subsection{Comparison to small-scale structure in other simulations}\label{sec:comparesims}

{While the idea of cool gas clumps forming in the halos of galaxies is not new (see, e.g., \citealt{KeresHernquist2009}), the } ability of cosmological zoom simulations to reach sub-kpc scales in the diffuse circumgalactic medium is a relatively recent development. Simulations of isolated galaxies or ``idealized'' portions of galaxies reached these scales first, owing to the efficiency gained from not following a large box, and at the cost of lacking a ``live halo'' \citep[e.g.,][]{Schneider2018, Abruzzo2023,Armillotta2017,Kopenhafer2023,Fielding2017}. This style of simulation demonstrates that there are  decisive theoretical reasons to pursue small-scale structures as tracers of accretion and feedback. 
Such simulations can reach sub-pc resolution and derive survivability criteria for structures on such scales, but lack the cosmological context of the cloud creation. 
Creation and survival of cool gas clumps at such small scales is heavily affected by local physical processes such as turbulent mixing and is a matter of open debate \citep[see, e.g.,][]{Abruzzo2023,Fielding2017}.
Reaching these scales is more challenging for cosmological zooms, owing to the need to carry the galaxy's cosmological environment as well. Numerical techniques have only recently matured to make such simulations feasible \citep{vandeVoort2019,Suresh2019,Hummels2019, Ramesh2024,Rey2004,Marra2024}. FOGGIE's production runs reach AMR cell sizes of 274 pc at $z = 0$ and 137 proper pc in the snapshots we analyze here. In these runs the median cell mass in the diffuse CGM is $\sim 44$ \Msun\ \citep{Peeples2019,Lochhaas2023}. While this is currently the state-of-the-art, there are already indications of structure smaller even than these scales. Other zooms have reached qualitatively similar resolution with AMR, SPH, or moving mesh techniques, so we can qualitatively compare their results.

{\citet{Hummels2019} have run simulations with forced refinement in the CGM and the same initial conditions as FOGGIE. 
They also analyzed clumps in their $z$ = 1 simulation output with the same clump identification method, but with a selection on \hi\ gas instead of the general gas density. They focused on characterizing the amounts and sizes of clumps with varying simulation resolution and they find a linear trend between cloud size and resolution with no sign of convergence. 
Furthermore, the number of clouds increase exponentially with increasing resolution, as the increased resolution allows more gas to cool and thus creates more cool clumpy gas overall.
Even though their selection was on \hi\ gas only, the result is applicable to our case as well, illustrating that with higher resolution we would find even smaller clouds.}

Recently, \citet{Ramesh2024} have analyzed the properties of cold clouds in the CGM of the GIBLE simulations. The GIBLE simulations are derived from the Illustris simulations, are run with AREPO, and use the same physical prescriptions inside galactic disks but with a high-resolution zoom region imposed in the CGM, where individual cells are $\sim 10^3$ \Msun\ and so clouds are resolved at $\gtrsim 10^4$ \Msun. GIBLE's findings are somewhat different from our results. While we find a striking maximum in the mass distribution of our clumps (Figure~\ref{fig:clump_hists}, Section~\ref{sec:global}), they find a logarithmic distribution with a sharp increase towards, followed by a cutoff at, lower masses. There are several potential reasons for this difference. First, \citeauthor{Ramesh2024} select only structures that are $\leq 10^{4.5}$\,K and comprised of $\geq 10$ cells. As a reminder, we only require our clumps to be local overdensities comprised of $\geq 20$ cells and our selection is agnostic with respect to the temperature of the gas. As can be seen in Figure~\ref{fig:hexplot3}, 30\% of our clumps are hotter than $10^{4.5}$\,K and 29\% of our clumps have masses $\ll 10^{4.5}$\,\Msun---the lower-mass cutoff of the clump distribution in \citeauthor{Ramesh2024}'s highest resolution simulation, where the individual cell mass in the CGM is $1.8\times 10^3$\,\Msun. For a minimum clump size of ten cells, this sets their minimum clump mass at $\sim 10^4$\,\Msun, much as our minimum number of cells sets a clear cutoff in our clump sphericalized radius measurement in Figure~\ref{fig:clump_hists}. However, since the FOGGIE refinement scheme does not impose a lower limit mass per cell (see also \citealp{Lochhaas2023}), resolved clumps at masses $\ll 10^{4}$\,\Msun\ are able to form. 
{Another factor for differences between these studies is the implemented feedback. 
While the FOGGIE simulations are run with stellar feedback only, GIBLE was run including both stellar and AGN feedback, which may have an impact on cloud formation in the CGM. For future CGM clump studies, a quantification of the impact of feedback on clump formation could be considered.
Finally, GIBLE uses magnetohydrodynamics (MHD), which has been shown to enable long-term survival of clouds \citep{McCourt2015}. 
FOGGIE simulations are run without MHD and therefore the overall clump population -- even if selected under the same criteria -- is expected to show some differences.}
Going forward, we believe that physical properties (such as mass and size functions) of uniformly-defined circumgalactic clumps will provide a useful additional metric for systematically comparing cosmological simulations \citep[e.g.,][]{Fielding2020b}.

\subsection{Comparison to small-scale structure in observations}\label{sec:compareobs}

FOGGIE's ability to resolve gas sub-structure at the kiloparsec scale in the circumgalactic medium suggests that we can now compare to observations that address the CGM at small physical scales. There are two broad categories of relevant observations to consider: (1) the ``high-velocity'' sky surrounding the Milky Way, where many unique constraints are available with a singular viewing geometry, and (2) external observations of QSO-absorber sightlines, where fewer physical constraints are available but they are taken from a complementary viewpoint. We address these two sets of observations in Sections \ref{sec:hvcs} and \ref{sec:multiple}, respectively.

\subsubsection{The Milky Way and Andromeda}\label{sec:hvcs}

\begin{figure*}[!t]
    \centering
    \includegraphics[width=\textwidth]{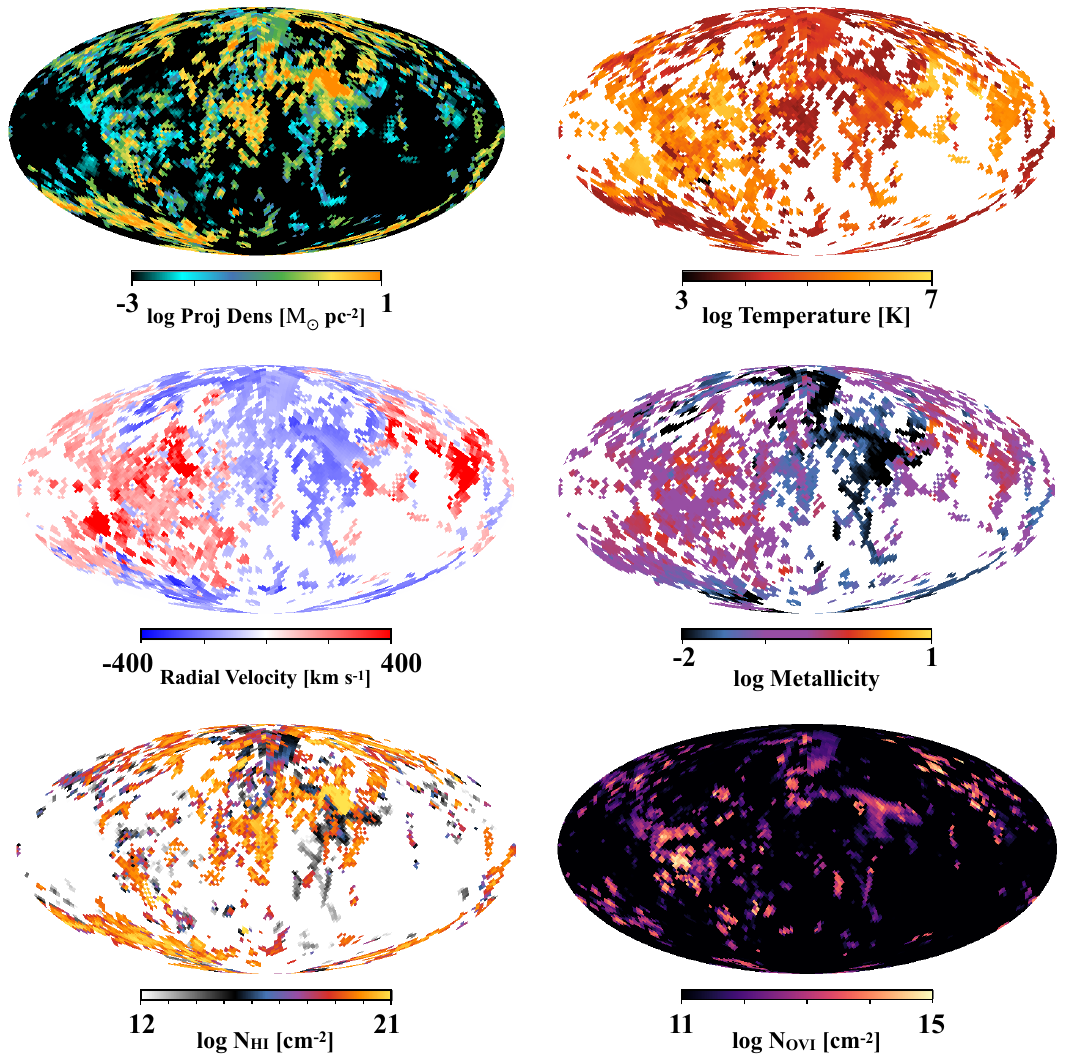}
    \caption{Six views of Tempest's clumpy CGM from inside its disk. From upper left: projected density, temperature, radial velocity, metallicity, \hi\ column density, and \ovi\ column density. Projected density and the two column densities are sums over all cells in projection, while the other quantities are mass-weighted mean values at each place on the map. This view has been filtered to include only clumps, not the ambient halo gas between them, and only for clumps with $|v| > 100$ km s$^{-1}$, to omit the ISM and mimic the Milky Way's high velocity clouds.}
    \label{fig:clumps_hvcs}
\end{figure*}

Up until now we have analyzed the clumps by considering their individual properties and the distribution of those properties across the population of clumps. We can also visualize the population of clumps in a different way, viewed as if we are inside the FOGGIE simulation, looking out. The population of small clumps we find in the highly resolved FOGGIE simulations cluster around $\sim 1$ kpc in size, and within $\lesssim 80$ kpc of their host galaxies. These geometric properties translate to angular diameters of 0.5--2$^{\circ}$ if we imagine viewing these clumps from inside the simulated galaxy.
Figure~\ref{fig:clumps_hvcs} shows six views of Tempest from the inside, viewed from the center of mass (not, as for the Milky Way, from 8 kpc outside the center; see also \citealp{Zheng2020}). These Aitoff projections show projected density and mass-weighted average temperature in the top row, mass-weighted average radial velocity and metallicity in the middle row, and projected
column densities for \ion{H}{1} and \ion{O}{6} in the bottom row. These renders contain {\it only} cells that are identified as being in clumps, so diffuse gas in between clumps is not shown. Additionally, only the clumps with radial velocities greater than 100 \kms\ (outflow) or less than $-100$ \kms\ (inflow) are shown. We note that, similar to the real Milky Way, this velocity selection excludes the ISM in the central galaxy but not necessarily the ISM in all of the satellite galaxies.

As in the observed Milky Way sky, the ``high-velocity'' projected gaseous halo of Tempest features both large ($>30^{\circ}$) and small ($<1^{\circ}$) \ion{H}{1} complexes. The \ion{H}{1} columns range from Ly$\alpha$ forest levels ($\lesssim 10^{15}$ cm$^{-2}$) to DLA strength ($>10^{20}$ cm$^{-2}$), much like is seen around the Milky Way \citep{fox2004, french2021}. Even though the absolute DLA covering fractions may be different in our simulated halo at cosmic noon compared to the Milky Way, this map can qualitatively be compared to the all-sky \hi\ map from the HI3PI survey \citep{Westmeier2018}. In the FOGGIE maps there is a clear bifurcation between cooler ``inflow'' regions and hotter ``outflow'' regions (see also in Figures \ref{fig:hexplot1} and \ref{fig:hexplot2}), with the latter dominating nearer the projected minor axes, as expected from bi-conical outflows that emerge perpendicular to the disk. The overall picture that emerges from this comparison is that FOGGIE's high-resolution treatment of CGM gas, based on forced and cooling refinement, naturally recovers a distribution of halo gas that resembles the Milky Way---even down to the small CHVCs. The median angular area of all clumps in this projection is 2.8 square degrees. 
We refrain from diving into a more quantitative comparison to the Milky Way HVC sky since in this work we are studying clumps at z=1 only, where the physical conditions of the CGM are different to z=0.
The Milky Way and our simulated halos are all relatively settled at z=0, whereas at z=1 they are only starting to settle and still have a lot more structure in their CGM than at later times. 
Moreover, their disks are different, as at higher redshift the disks will naturally be smaller than at z=0. 
{Crucially, at z=0 these halos contain only little \hi\ bearing cool gas in the halos \citep{Zheng2020}, causing a lack of clumpy structures to compare to HVCs.}
For a more quantitative comparison of the FOGGIE halos to the Milky Way, we refer the reader to \citet{Zheng2020}.
Nevertheless, we deem this qualitative comparison of simulated clumps and observed HVCs insightful, as we find remarkable similarities.

FOGGIE qualitatively agrees with observations of the Milky Way in a basic respect: the simulations recover a large population of kiloparsec-scale clouds that project degree-scale angular sizes when viewed from the inside. These sizes are consistent with the CHVCs in the Milky Way halo \citep{Putman2012,Westmeier2018}, under the assumption that those CHVCs lie at similar distances (sizes 1--2 kpc for $d = 50$ kpc). If simulations did not recover hundreds to thousands of small clouds at this scale, they would fail a basic observational test. Furthermore, the finding that small clumps lie preferentially nearer the disk, within 20--40 kpc, with low ions favoring smaller radii, is consistent with the finding from Project AMIGA \citep{Lehner2020} that low ions are found preferentially near M31's disk. 
The trend of finding more low ionization gas components (clumps) closer to the disk has also recently been found from observations \citep{Sameer2024}, and is in agreement with our simulated clumps. 

As for the origin of HVCs in simulations, unfortunately the current FOGGIE simulation runs lack the robust gas tracking capabilities needed to provide an answer.
However, \citet{Lucchini2024} have studied the origin of HVCs in the IllustrisTNG simulations and found them to have diverse origins, with the large majority of 62\%\ condensing out of the CGM and the rest coming from disk or satellite material.
This is qualitatively in agreement with our findings where HVCs seem to be a natural subset of the clumps we find in our halos, where most of them are in the CGM and only smaller fractions belong to satellites or disks (see Figure \ref{fig:hexplot1}).

\subsubsection{Small-scale structure from other probes of external galaxies}\label{sec:multiple}
Another line of evidence for small scale structures in the CGM comes from quasar absorption line studies that use closely-spaced pencil beam sightlines or extended sources as the background. 
In a pioneering study, \citet{Rauch2001} examined \civ\ detections in the spectra of closely-spaced multiple images of lensed quasars. 
They found that the variation in line strength was minimal (a few percent) at physical scales of $\sim 100$ pc, but increased rapidly to 40\% variations at 100--1000 pc and significantly higher variations ($\sim 80$\%) above 1 kpc). 
In a followup study \citet{Rauch2002} found variations in \mgii\ strength at the 200--400 pc scale in multiply lensed QSOs. 
Another, more indirect,  measure of \mgii\ absorber size comes from model-based indications for small, $\simeq 30$ pc cloud absorbers from fitting the global distribution of \mgii\ absorbers in the SDSS sample \citep[e.g.,][]{Lan2017}. 
Finally, there are indications from resolved, gravitationally lensed systems, for large (10$\times$) variations in total \hi\ column density over $\sim$kpc scales \citep[e.g.][]{Bordoloi2022Nature}. 
{\citet{Rubin2018} compiled a number of absorber variations against lensed quasars from the literature for both high and low ions and find that higher ions show coherence over larger scales, while lower ions such as \mgii\ already vary strongly on kpc scales.
\citet{Augustin2021} later added to this compilation of \mgii\ absorbers towards lensed quasars and explored the expected \mgii\ coherence in FOGGIE simulations, finding that both in simulations and observations there can be very strong variation on kpc scales and below.}
Particularly, \citet{Afruni2023} studied the \mgii\ coherence length toward a giant lensed arc and constrained \mgii\ cloud sizes to be between 1.4 and 7.8 kpc. 
As we have seen from our study, \mgii\ as a low ionization line is typically well confined within the simulated clumps.
Of course, a direct comparison with the simulated clumps is difficult since coherence length does not necessarily translate into cloud size, but both are on $\sim$ kpc scales.
{However, we see some consistency that the \mgii\ absorber variations agree with our clump sizes on kpc scales.}
Taken altogether, these observations provide multiple lines of evidence for a CGM that is structured at and below the kiloparsec scale when viewed by multiple tracers from \hi\ to \civ, but not much on scales $\lesssim 100$\,pc.

\subsection{Insights about Small-scale structure}\label{sec:insights}

Overall, the highly structured nature of the CGM as rendered by FOGGIE suggests a shift in our thinking away from ``dense versus diffuse'' gas when trying to understand CGM gas structure from absorber variation. A more apt line of thought would be ``structured vs homogeneous'' gas. This is because ``diffuse'' can mean homogeneous around any given overdensity even though  ``diffuse'' gas will have different densities in different regions.

\textit{Size limit of CGM clumps:}
A caveat in any simulation is the resolution limit at which clouds are being investigated. 
The higher the resolution, the more clouds start to shatter and become smaller. 
We can see in Figure \ref{fig:clump_hists} the clear peak towards the resolution limit in the clump size distribution.
If clumps do break up into smaller pieces, would they still be as longlived as we predict, or would they rather enter a cycle of destruction and re-condensation?
Future works on even higher resolved CGM clumps, where individual clouds would be treated like a wind-tunnel simulation but with the cosmological context, may have an answer.
{An additional theoretical consideration is that there is likely a natural physical limit to the minimum size of a ``clump'' as a discrete structure -- i.e., a limit imposed by the interaction mean free path of the particles that comprise that clump.  An overdense object that is smaller than several mean free paths in diameter will quickly disappear.  But, this raises a question -- what is the appropriate mean free path in a clump?  For our smallest clumps (Using M$_{\mathrm{clump}} \sim 10^4$~M$_\odot$, R$_{\mathrm{clump}} \sim 250$~pc from Figure~\ref{fig:hexplot1}) one finds a typical number density of $\simeq 0.01$~cm$^{-3}$. 
Neutral atoms interact only by collisions, which means that for a neutral hydrogen atom with a collisional cross section having a radius of approximately the Bohr radius one gets a mean free path of multiple kiloparsecs -- a scale vastly exceeding the measured clump size in these calculations.  
In Figure \ref{fig:hexplot1} we have seen that most clumps are at least partially ionized, however, and in the absence of magnetic fields the primary interaction scale is that of Coulomb interaction, with $\lambda_c \simeq 1.25 \left(\frac{\mathrm{n}}{0.01 \mathrm{cm}^{-3}}\right)^{-1}\left(\frac{\mathrm{T}}{10^4 \mathrm{K}}\right)^{2}$ au for a plasma comprised of ionized hydrogen \citep{beresnyak20232023}.  This is a scale that is much smaller than the size of the cloud, though we note that for plasma outside of a clump, with temperatures and densities appropriate for the ambient CGM in a Milky Way-mass galaxy, this distance is in the $1-100$~pc range.  Magnetic fields are likely to be present in the CGM, and charged particles will effectively be bound to these fields.  In the case of a $\sim 1~\mu$G magnetic field and $T_e = T_i \simeq 10^4$~K this results in electron and ion gyroradii on the scale of $\simeq 10^6$ and $10^8$~cm for a hydrogen plasma, respectively -- even smaller than the Coulomb mean free path.  Overall, this suggests that neutral atoms may exhibit behavior that is quite different than charged particles, which is something that cannot be captured in simulations that model the circumgalactic medium using the single-fluid approximation (which is to say, any cosmological simulation).  This was briefly noted as a potential issue in \citet{McCourt2018}, but the consequences remain unexplored and are an avenue for future exploration.}

In this context, \citep{vandeVoort2021} have explored the impact of magnetic fields in high resolution CGM simulations. 
They find that the presence of magnetic fields has a noticeable impact on the formation of small scale structure, enhancing the clumpiness of the CGM even further.
In particular, they note that the larger scatter in metallicity implies that metals do not mix as well as in the absence of magnetic fields, resulting in pockets of low metallicity dense gas.  This could potentially enhance the effect of low metallicity dense clouds, which we find in our simulation, even more.
\citep{Ramesh2024MF} have studied the impact of cool CGM gas clumps on the draping of magnetic field lines and found a diverse variety of field line arrangements around the clumps, suggesting that strongly draped configurations around CGM clumps are rare but that some properties such as cloud overdensity and relative velocity can have an impact on magnetic field topology.

{Besides magnetic fields, cosmic rays (CRs) can also have an impact on small-scale structure in the CGM.
\citet{Butsky2020} found that CRs can prevent cool gas formation from thermal instabilities.
\citet{Ji2020} found that CRs have an impact on overall cloud density and size, as they allow for larger, more diffuse gas clouds out of thermal pressure equilibrium with their surroundings, which can also be more longlived.
\citet{Weber2025} investigated different CR transport modes and found that the implementation of CR transport has a significant effect on how CRs affect CGM clumps.
It seems therefore that the exact effect of CRs on CGM small-scale structure is still in debate, but it is clear that inclusion of CRs can have drastic impact on the clump population overall.}

\textit{Impact of CGM clumps on observations:}
We have also seen that the cool gas content in particular increases significantly with increasing resolution \citep{Peeples2019,vandeVoort2019,Hummels2019,Corlies2020}, raising the question as to what resolutions this trend holds given that not all of the gas in the real universe is cold.
On the other hand, we have seen that our clumps are not necessarily cool.
Some of them trace $10^{6}$ K gas and they can have significant internal temperature gradients. 
We have seen in the clump projections (Figures \ref{fig:individual_clump_physical1} and \ref{fig:individual_clump_physical2}) that particularly the OVI-tracing gas sits preferentially at the edges of the clumpy structures. 
With future space-based missions with OVI mapping capabilities, such as Aspera \citep{Chung2021} and the Habitable Worlds Observatory \citep{France2024}, further investigation of what physical structures underlie the observed OVI maps will be necessary.
More generally, we may have to rethink the traditional view of the hot halo that is filled with cool clouds into a more complex picture of a multiphase diffuse medium with internal multiphase gas structures.
A more complex picture like this will also challenge mock observables drawn from idealized models and potentially bring us closer to bridging the gap between observed and simulated small-scale structures.
Tools such as CloudFlex \citep{Hummels2024}, in combination with clumps found in a cosmological context as presented in this paper, may help us in future works to create more meaningful mock observables.

\section{Conclusions}
\label{sec:conclusions}

We have analyzed the properties of natural overdensities in the CGM and their implications for the observable properties of Milky Way-like galaxies. We identified clumps using a purely theoretical, density-based selection of simply-connected local overdensities regardless of their internal properties or shape. We then studied the internal properties of clumps (such as temperature or metallicity) and investigated global trends with respect to the host galaxy. Performing this analysis on AMR simulations like FOGGIE brings several key benefits: (1) we reach clump sizes down to $\sim 250$ pc (as measured by sphericalized radius; one FOGGIE AMR cell at $z = 1$ is a cube 137 proper pc on a side), (2) computational cells are defined in space, not by mass, so that clumps and their environments can be assessed on a consistent physical scale (this contrasts with Lagrangian codes, in which hot gas is followed at a coarser spatial resolution), and (3) clumps can be detected as lower overdensity objects even when they are not cold and dense.

Our analysis leads us to several key conclusions: 
\begin{enumerate}

\item The CGM is highly inhomogeneous and structured, with large density fluctuations within single halos and substantial variation from halo to halo. 

\item Density-bounded clumps come in all shapes and sizes with numbers that continue to increase to the adopted resolution limit, suggesting that still smaller and more numerous clumps lie below our size resolution limit.

\item These clumps have higher densities than their immediate environments, by definition, but a subset are larger, more diffuse, and warmer than the majority of the population. Most clumps are generally in pressure equilibrium with their immediate surrounding, except at $M \gtrsim 10^6 M_{\odot}$ where small clumps are under-pressured and large ones are over-pressured.

\item Even though species with different ionization potentials are often observed at the same velocity, they need not probe the same volume  ranges in the CGM. Even clumps at the small end of the size distribution can show internal ionization structure, suggesting that closely-spaced kinematic components of different ionization states can nevertheless arise in a single, relatively small structure. 

\item In a projected view or in observational scenarios with multiple absorbers, we should expect large variations of \hi\ and low ion gas on $\sim$\,kpc scales, {similar to what is found for absorber variations at $z = 1$ towards lensed background sources (see e.g. \citet{Augustin2021,Afruni2023}).} Higher ions should appear more extended and more homogeneously distributed with clumps up to $\sim 10$ kpc in size and a larger contribution from the more diffuse inter-clump gas. This contrast should be more pronounced in the inner regions of the CGM, where our clumps are concentrated. This finding is compatible with the results of Project AMIGA. 

\item We find that all CGM clouds, independent of their location, size, mass, or ionization state, are metal-poor with respect to their immediate environment. This means that metals are preferentially embedded in the diffuse halo, rather than the dense clouds. This may have some impact on line-of-sight metallicity measurements, as the metals and the \hi-bearing gas may probe different parts of the halo.

\item We investigated the survivability of the FOGGIE clumps using the criterion described in \citet{Abruzzo2023} and find that clumps that are larger than 0.5 kpc should generally survive. A small fraction of the $<0.5$\,kpc clumps may shatter. This longevity is mainly due to the relatively small velocity shear between clouds and their immediate environments. This is different from the well-studied wind tunnel simulations (e.g. \citealt{Schneider2018,Abruzzo2023}) where the shear velocities can be as high as $1{,}000$ \kms.  Clouds in such simulations are, therefore, less likely to survive compared to FOGGIE clumps where we have the full cosmological environment context.

\end{enumerate}

We find qualitatively broad agreement between FOGGIE's small-scale structure and the CGM as observed in the Milky Way, M31, and external galaxies where relevant data is available. However, we caution against over-interpretation of these similarities or the notable differences. While the broad agreement about the widespread presence of small-scale structure is encouraging, the simulated and real HVC sky are not congruent in many respects.
In the Milky Way, we see a snapshot of a real galaxy with a certain merger history, at a particular time in its history, and the current set of satellites, the distributions of size, mass, metal content, and kinematics are not necessarily generic properties that are maintained over time. Therefore, it may not be possible to match simulations with the specifics of Milky Way observations. Hence, while FOGGIE does not match in detail the exact angular size distribution, metallicities, or kinematics of the Milky Way halo, it is still instructive that they agree in broad strokes.
We also do not know from observations how much variation in the small-scale structure of the CGM occurs naturally from galaxy to galaxy. The galaxy-to-galaxy variation is hinted at by comparisons between the Milky Way and M31, and the scales of spatial and kinematic variation are constrained by the multi-QSO and lensing data at high-redshift. Much more work, beyond the scope of this initial paper, is required to show how the time-varying history of small-scale structure evolves in simulations and to quantitatively compare to the Milky Way's current state, and that of other galaxies over time. 

\section*{Acknowledgements}
We thank Andrew Fox and Francie Cashman for helpful discussions about the Galactic HVCs. 
RA and AA thank Beena Meena for organizing paper writing sessions that helped finish this manuscript.
RA, CL, AA, and MSP were supported for this work in part by NASA via an Astrophysics Theory Program grant 80NSSC18K1105. 
RA and CL also acknowledge financial support from the STScI Director’s Discretionary Research Fund (DDRF). 
RA acknowledges funding by the European Research Council through ERC-AdG SPECMAP-CGM, GA 101020943. 
RA's efforts for this work were additionally supported by HST AR \#15012 and HST GO \#16730. 
BWO acknowledges support from NSF grants \#1908109 and \#2106575 and NASA ATP grants NNX15AP39G and 80NSSC18K1105. 
JT and ACW acknowledge support from the Roman Space Telescope Milky Way Science Investigation Team. 
AA has also been supported by NSF-AST 1910414 and HST AR \#16151. Support for CL was provided by NASA through the NASA Hubble Fellowship grant \#HST-HF2-51538.001-A, awarded by the Space Telescope Science Institute, which is operated by the Association of Universities for Research in Astronomy, Inc., for NASA, under contract NAS5-26555.

Resources supporting this work were provided by the NASA High-End Computing (HEC) Program through the NASA Advanced Supercomputing (NAS) Division at Ames Research Center and were sponsored by NASA's Science Mission Directorate; we are grateful for the superb user-support provided by NAS. 

\texttt{Enzo} \citep{Bryan2014,BrummelSmith2019} and \texttt{yt} \citep{Turk2011} are developed by a large number of independent researchers from numerous institutions around the world.  This research made use of Astropy (http://www.astropy.org), a community-developed core Python package for Astronomy \citep{Astropy1, Astropy2, astropy2022}. Their commitment to open science has helped make this work possible. The python packages {\sc matplotlib} \citep{hunter2007}, {\sc numpy} \citep{walt2011numpy}, {\sc rockstar} \citep{Behroozi2013a}, and {\sc tangos} \citep{pontzen2018} were also used in parts of this analysis or in products used by this paper.

\vspace{5mm}

\appendix

\section{Clump Finding Parameters}

\begin{figure}
    \centering
    \includegraphics[width=0.5\linewidth]{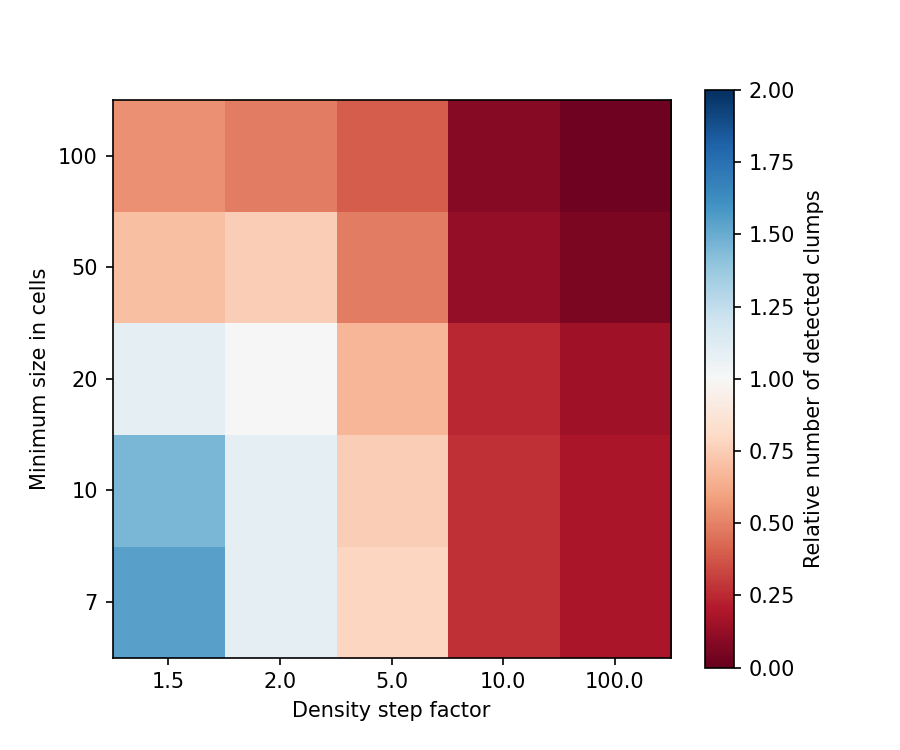}
    \caption{{Effect of choice of clump finding initial parameters. Our fiducial choice is 20 cells as the minimum size and a factor 2 increase between density steps. We ran the algorithm on a small patch of CGM in one halo and illustrate how many more or fewer clumps we would find for a different choice of parameters.}}
    \label{fig:clumpfinderparameters}
\end{figure}

{Here we briefly outline the effect of the choice of parameters for the clump finding process.
For the clumps presented in this paper, we chose a minimum clump size of 20 connected cells to make sure the clumps have some internal resolution and we do not end up with artifacts that are just one cell wide, in any given spatial direction.
For the density threshold, we chose a factor of 2 at every iteration, which is a compromise between run time and sampling the density ranges somewhat smoothly.
Of course, choosing a different size criterion and a different step in density would change the number of total detected clumps in the halo, but is not expected to affect the extracted physical properties of the clumps. 
To quantify the effect of different parameters, we ran the clump finding algorithm on a random patch of CGM in one halo. 
Figure \ref{fig:clumpfinderparameters} shows the fractional change in the number of detected clumps relative to our fiducial parameter choices.
From this test we can see that for the 20 cell criterion, lowering the density step does not significantly increase the number of detected clumps. 
Similarly, if we keep the same density step but lower the minimum size criterion, not significantly more clumps are found. 
Only a combined change of density step factor and minimum size criterion can increase the number of detected clumps significantly. However, this is due to many small artifacts that masquerade as clumps, as described above.}

\bibliography{sample631}{}
\bibliographystyle{aasjournal}

\end{document}